\documentclass[revtex4]{emulateapj}
\usepackage{epsf}
\usepackage{natbib}
\usepackage{graphicx}
\usepackage{float}
\usepackage{afterpage}
\usepackage{amssymb}
\usepackage{enumitem}
\usepackage{bm,bbm,amssymb, amsmath}
\usepackage{enumerate}
\usepackage{url}
\usepackage[section]{placeins}

\usepackage{mathrsfs}
\usepackage{ulem}

\usepackage{hyperref}
\usepackage[usenames]{color}

\begin{document}
\title{Thermal damping of Weak Magnetosonic Turbulence in the Interstellar Medium}
\author{Kedron Silsbee$^1$, Alexei V. Ivlev$^1$, Munan Gong$^1$}
\email[e-mail:~]{ksilsbee@mpe.mpg.de} \email[e-mail:~]{ivlev@mpe.mpg.de} \affiliation{$^1$Max-Planck-Institut f\"ur
Extraterrestrische Physik, 85748 Garching, Germany }

\begin{abstract}
We present a generic mechanism for the thermal damping of compressive waves in the interstellar medium (ISM), occurring due
to radiative cooling. We solve for the dispersion relation of magnetosonic waves in a two-fluid (ion-neutral) system in
which density- and temperature-dependent heating and cooling mechanisms are present.  We use this dispersion relation, in
addition to an analytic approximation for the nonlinear turbulent cascade, to model dissipation of weak magnetosonic
turbulence. We show that in some ISM conditions, the cutoff wavelength for magnetosonic turbulence becomes tens to
hundreds of times larger when the thermal damping is added to the regular ion-neutral damping.  We also run numerical
simulations which confirm that this effect has a dramatic impact on cascade of compressive wave modes.
\end{abstract}

\section{Introduction}
\label{Introduction}

Turbulence is ubiquitous in the interstellar medium (ISM), and is extended over an extremely wide range of scales
\citep{Chepurnov10, Armstrong95, Minter96, Marchal2021}.  Turbulence on large scales plays an essential role in the theory of star
formation \citep{McKee07}.  The origin of this turbulence is debated \citep{MacLow04, Glazebrook13, Krumholz16}.  Two main
classes of driving mechanisms are gravitational instabilities and supernova blast waves.  Both input energy at scales of
parsecs or larger.  The turbulent cascade then brings this energy to small scales, generating density and velocity structure
which profoundly impacts the star formation process.

The ISM is composed of gas with a wide range of ionization degrees, ranging from almost completely ionized near massive
stars \citep{Stromgren39} to lower than $10^{-8}$ in dense molecular cloud cores \citep{Caselli02}.  It is currently thought
that MHD turbulence in partially ionized gas in the ISM is damped primarily by ion-neutral friction \citep{Kulsrud69,
Balsara96}.  \citet{Xu16} provide an
analytic calculation of the scales at which different wave modes in the gas are damped. \citet{Li08} study the line widths of both charged and neutral species in a molecular cloud, showing that
the ion widths are systematically narrower. 
They suggest that this occurs because the ion
motions are damped more strongly than the neutral motions at the scale of ion-neutral decoupling.   
Interestingly though, \citet{Pineda21} find in a different cloud that the ions have a systematically {\it higher} velocity dispersion than the neutrals, indicating the need for a more detailed analysis of the turbulent cascade at small scales.

Understanding the turbulent cascade at small scales is important for a number of astrophysical applications, particularly in
molecular clouds.  Turbulence at scales of an AU or less affects the transport of the sub-GeV cosmic rays (CRs) which determine the heating and ionization of molecular gas \citep{Yan04, Xu13, Xu16, Silsbee19}.  Turbulent eddies on small scales \citep{YLD04, Ormel07} are also
likely the dominant source of relative motion between dust grains in many environments, thus determining the grain
coagulation \citep{Ossenkopf93, Hirashita09, Gong20, Silsbee20}.   

In this paper we show that compressive (magnetosonic) modes in weak MHD turbulence are strongly damped due to radiative energy
losses. This thermal damping can occur at much larger scales than those on which ion-neutral friction becomes important.
The damping mechanism is as follows. The steady-state gas temperature is determined by a balance of global heating and
cooling. As a compressive disturbance propagates, the gas undergoes adiabatic heating and cooling in response to the
pressure perturbation, while the global processes bring it back to an equilibrium state on some characteristic timescale,
which we call the {\it cooling time} $\tau_{\rm c}$.  If $\tau_{\rm c}$ is large compared to $\omega^{-1}$, then an adiabatic equation
of state is appropriate and thermal damping is negligible. If $\tau_{\rm c}$ is small, then the wave speed may be modified (by the
different equilibrium temperatures in the compressed and rarefied parts of the wave), but the wave amplitude also remains
constant.  However, if the cooling time is comparable to $\omega^{-1}$, then the restoring pressure force (which governs the
wave propagation) is reduced, thus damping the wave. This damping effect has been studied previously in the context of waves
in the solar atmosphere \citep[e.g.][]{Souffrin72, Mihalas83, Bunte94}.  \citet{Tilley11} derive a very general dispersion relation for two-fluid MHD, considering both cooling and self-gravity.  They point out that the cooling can increase the damping of certain wave modes in molecular clouds, but do not explore under which circumstances this will substantially affect the turbulent cascade.
In the present paper, we focus on the effect of such damping on the turbulent cascade in the ISM and explore the regimes under which its effect is significant.

In Section \ref{DRderiv}, we derive the dispersion relation for magnetosonic waves in a two-fluid ion-neutral plasma in the
presence of thermal damping. We plot the dispersion relation for realistic astrophysical
environments in Section \ref{behavior}, and in Section~\ref{sect:magnitude} discuss in detail the conditions required in order for the effect to be large.  In Section \ref{cascade}, we present a simplified model for the
turbulent cascade, and show that under reasonable assumptions, thermal damping may play a dominant role in damping
turbulence.  In Section \ref{simulation}, we describe the results of numerical simulations illustrating this effect. Several potential implications of thermal damping are outlined in Section \ref{sect:impl}, and our conclusions are presented in Section \ref{sect:conclusion}.

\section{Dispersion relation}
\label{DRderiv}

In this section we derive the dispersion relation for compressive MHD waves propagating in a medium in which there are gas
heating and cooling mechanisms. Assuming small perturbations of density, $\rho = \rho_0 + \delta \rho$, and temperature, $T
= T_0 + \delta T$, we can write a linearized energy conservation equation \citep[see, e.g.,][]{Landau60}, where the thermal
and magnetic parts of the energy remain decoupled.  The resulting equation for the thermal energy balance reads:
\begin{equation}
C_v \left(\rho_0 \frac{\partial \delta T}{\partial t} + T_0 \frac{\partial \delta \rho}{\partial t}\right) + C_p \rho_0T_0
\nabla \cdot {\bf v} = \delta q.
\label{eq1}
\end{equation}
Here, $\bf v$ is the fluid velocity, $t$ is time, $\rho \approx \rho_i + \rho_n$ is the total gas density, $C_v$ and $C_p$
are the gas heat capacities (per unit mass) at constant volume and pressure, and $q(\rho, T) \equiv \Lambda_{\rm h}(\rho, T) -
\Lambda_{\rm c}(\rho, T)$ is the net energy deposition (per unit volume) from all heating and cooling processes.  In addition, we
have the linearized continuity equation:
\begin{equation}
\frac{\partial \delta \rho}{\partial t} + \rho_0 \nabla \cdot {\bf v} = 0.
\label{continuity}
\end{equation}
Equations \eqref{eq1} and \eqref{continuity} yield the relation
\begin{equation}
\frac{\partial \delta T}{\partial t} - \left(\gamma - 1\right) \frac{T_0}{\rho_0} \frac{\partial \delta \rho}{\partial t} =
\frac{q'_\rho \delta \rho + q'_T \delta T}{\rho_0 C_v}\,,
\label{randomIntermediate}
\end{equation}
where $\gamma \equiv C_p/C_v$, $q'_\rho \equiv \partial q/\partial \rho$ and $q'_T \equiv \partial q/\partial T$. Now, let
us assume harmonic perturbations proportional to $\exp{(-i\omega t )}$.  Then, defining
\begin{equation}
 \tau_{\rm c} = \frac{C_v \rho_0}{-q'_T} \quad {\rm and} \quad \Phi = \frac{\rho_0q'_\rho}{T_0 q'_T}\,,
 \label{psiAndAlpha}
\end{equation}
we find from Equation \eqref{randomIntermediate},
\begin{equation}
\left(1 - \frac{1}{i\omega \tau_{\rm c}}\right) \frac{\delta T}{T_0} =  \left(\gamma-1 + \frac{\Phi}{i\omega \tau_{\rm c} }\right)
\frac{\delta \rho}{\rho_0}\,.
\label{lastEnergyEquation}
\end{equation}
We note that $\tau_{\rm c}$ is always positive because $q'_T$ must be negative to ensure thermal stability. We find in Section
\ref{sect:effect} that the thermal damping effect is only relevant for $k$ such that ions and neutrals experience many
collisions per $\omega^{-1}$.  This allows us to consider the plasma as one fluid, assuming the ions and neutrals to be
perfectly coupled, both thermally and mechanically.

In what follows, we introduce the effect of thermal damping into the standard framework of the two-fluid approach. For this,
we consider the separate motion of ions and neutrals, still assuming that ions and neutrals are perfectly thermally coupled.
Following Equations 1--3 in \citet{Soler13}, we then write a linearized equation of motion for ion-electron fluid:
\begin{equation}
\rho_i \frac{\partial {\bf v}_i}{\partial t} = - \nabla \delta p_i   + \frac{1}{4 \pi} \left(\nabla \times \delta {\bf B} \right)
\times {\bf B}_0 - \xi \left({\bf v}_i - {\bf v}_n\right).
\label{forceIon}
\end{equation}
Here ${\bf B}_0$ and $\delta {\bf B}$ are the mean magnetic field and its perturbation, ${\bf v}_i$ and ${\bf v}_n$ are the
ion-electron and neutral velocities, respectively, and $\delta p_i$ is the pressure perturbation of the ion-electron fluid.
$\delta {\bf B}$ obeys the linearized equation
\begin{equation}
\frac{\partial \delta {\bf B}}{\partial t} = \nabla \times ({\bf v}_i \times {\bf B}_0).
\label{eq:bdot}
\end{equation}
$\xi$ is a constant describing the magnitude of the ion-neutral friction, given by
\begin{equation}
\xi = \frac{\rho_i\rho_n \langle \sigma v \rangle_{in}}{m_i + m_n}\,,
\end{equation}
where $\langle \sigma v \rangle_{in} = 4 \pi a_0^2 \sqrt{(\alpha/a_0^3)I_{\rm H}/\mu}$ is expressed via the Bohr radius
$a_0$, the Rydberg energy $I_{\rm H}$, and $\mu = m_i m_n/(m_i + m_n)$.  The dimensionless polarizability of the neutral
species, $\alpha/a_0^3$, is 4.5 for atomic hydrogen and 5.52 for molecular hydrogen \citep{Raizer11}.

Similarly, the equation of motion for the neutrals reads
\begin{equation}
\rho_n \frac{\partial {\bf v}_n}{\partial t} = - \nabla \delta p_n - \xi \left({\bf v}_n - {\bf v}_i\right).
\label{forceNeutral}
\end{equation}
By writing $\delta p_{i, n} = n_{i,n} k_{\rm B} \delta T + k_{\rm B} T_0 \delta n_{i, n}$, where $k_{\rm B}$ is the Boltzmann constant, $n_i \approx 2\rho_{i}/m_{i}$ and $n_n
=\rho_n/m_n$, we obtain from Equation \eqref{lastEnergyEquation} for harmonic perturbations
\begin{equation}
\delta p_{i, n} = c_{i, n}^2 f(\omega) \delta \rho_{i, n}\,,
\label{eq12}
\end{equation}
where
\begin{equation}
c_i = \sqrt{\frac{2\gamma k_{\rm B} T_0}{m_i}}\,,
\label{eq:ci}
\end{equation}
\begin{equation}
c_n = \sqrt{\frac{\gamma k_{\rm B} T_0}{m_n}}\,,
\label{eq:cn}
\end{equation}
are the sound speeds of ions and neutrals, respectively,
\begin{equation}
f(\omega) = \frac{i\omega \tau_{\rm c} - \chi}{i \omega \tau_{\rm c} - 1}\,,
\label{eq:fdef}
\end{equation}
and
\begin{equation}
\chi = \frac{1-\Phi}{\gamma}\,.
\label{eq:chi}
\end{equation}
We note that $\chi > 0$ is the requirement for a stable equilibrium to exist. As shown in
Appendix~\ref{app:modifiedDamping}, the imaginary part of the frequency in the long-wavelength regime becomes positive when
$\chi > 1$.  This would imply a thermal instability.  In principle, this could be realized if the concentration of a
dominant coolant were strongly suppressed at higher density.  We do not consider such a hypothetical situation in this
paper, and thus assume $0<\chi<1$.

Combining Equation \eqref{eq12} with Equation \eqref{continuity}, we obtain
\begin{equation}
i\omega \delta p_{i, n} = c_{i, n}^2 \rho_{i, n} f(\omega) \nabla \cdot {\bf v}_{i, n}\,.
\label{eq:EOS}
\end{equation}
Note that we are are interested in magnetosonic perturbations, as non-compressive Alfv\'{e}nic perturbations are not
radiatively damped. Hence, following \citet{Soler13}, we take the time derivative of the divergence of Equation
\eqref{forceIon}.  Using Equations \eqref{eq:bdot} and \eqref{eq:EOS} and writing $\Delta_{i,n} = \nabla \cdot {\bf  v}_{i,n}$ for brevity, we find
\begin{align}
& -\omega^2 \rho_i \Delta_i = c_i^2 \rho_i f(\omega) \nabla^2 \Delta_i
\nonumber \\
&+ \frac{1}{4\pi} \nabla \cdot \left\{\left[\nabla \times \nabla \times ({\bf v}_i \times {\bf B}_0)\right] \times
{\bf B}_0\right\} + i\omega \xi \left(\Delta_i - \Delta_n\right).
\label{doubleCurlEq}
\end{align}
For ${\bf B}_0 = B_0 \hat {\bf z}$, some manipulation of the middle term in Equation \eqref{doubleCurlEq} allows us to write
\begin{align}
&-\omega^2 \rho_i \Delta_i = c_i^2 \rho_i f(\omega) \nabla^2 \Delta_i
\nonumber \\
& +\frac{B_0^2}{4\pi} \nabla^2\left(\Delta_i - \Delta_i^\parallel \right) +i\omega \xi \left(\Delta_i - \Delta_n\right),
\end{align}
where $\Delta_i^\parallel \equiv dv_i^z/dz$ is the parallel component of the divergence. Taking the time derivative of the parallel divergence of Equation \eqref{forceIon} yields
\begin{equation}
-\omega^2 \rho_i \Delta_i^\parallel = c_i^2 \rho_i f(\omega) \frac{\partial^2 \Delta_i}{\partial z^2} +i\omega \xi
\left(\Delta_i^\parallel - \Delta_n^\parallel \right).
\end{equation}
A complementary pair of equations for $\Delta_n$ and $\Delta_n^\parallel$ are derived from Equations \eqref{forceNeutral} and \eqref{eq:EOS}:
\begin{equation}
    -\omega^2 \rho_n \Delta_n = \rho_n c_n^2 f(\omega) \nabla^2 \Delta_n + i \omega \xi(\Delta_n - \Delta_i),
\end{equation}
\begin{equation}
   -\omega^2 \rho_n \Delta_n^\parallel = c_n^2 \rho_n f(\omega) \frac{\partial^2 \Delta_n}{\partial z^2} + i \omega \xi (\Delta_n^\parallel - \Delta_i^\parallel).
\end{equation}

Defining the Alfv{\'e}n speed,
\begin{equation}
c_{\rm A} = \frac{B_0}{\sqrt{4\pi \rho_i}}\,,
\label{eq:cA}
\end{equation}
the ion-neutral and neutral-ion momentum transfer frequencies,
\begin{equation}
\nu_{in} = \frac{\xi}{\rho_i} \quad {\rm and} \quad \nu_{ni} = \frac{\xi}{\rho_n}\,,
\label{eq:nus}
\end{equation}
and assuming all perturbed quantities to be proportional to $\exp(-i \omega t + i{\bf k} \cdot {\bf r})$, we can write the
following 4 coupled equations
\begin{equation}
\left[\omega^2 - k^2c_i^2 f(\omega)\right] \Delta_i + i\omega \nu_{in}(\Delta_i - \Delta_n) = k^2 c_{\rm A}^2 \left(\Delta_i -
\Delta_i^\parallel\right),
\label{eq20}
\end{equation}
\begin{equation}
\left[\omega^2 - k^2 c_n^2 f(\omega)\right]\Delta_n + i\omega \nu_{ni} (\Delta_n - \Delta_i) = 0,
\label{eq20a}
\end{equation}
\begin{equation}
\omega^2 \Delta_{i, n}^\parallel  -k_z^2 c_{i, n}^2 f(\omega) \Delta_{i, n}  \pm i \omega \nu_{in, ni}
\left(\Delta_i^\parallel - \Delta_n^\parallel \right) = 0.
\label{eq21}
\end{equation}
Equations \eqref{eq20} - \eqref{eq21} yield a 9$^{\rm th}$ order polynomial dispersion relation describing the modes present in a partially ionized medium with thermal damping.

\section{Magnetosonic modes in astrophysical environments}
\label{behavior}

\begin{figure*}[htp]
\centering
\includegraphics[width = \textwidth]{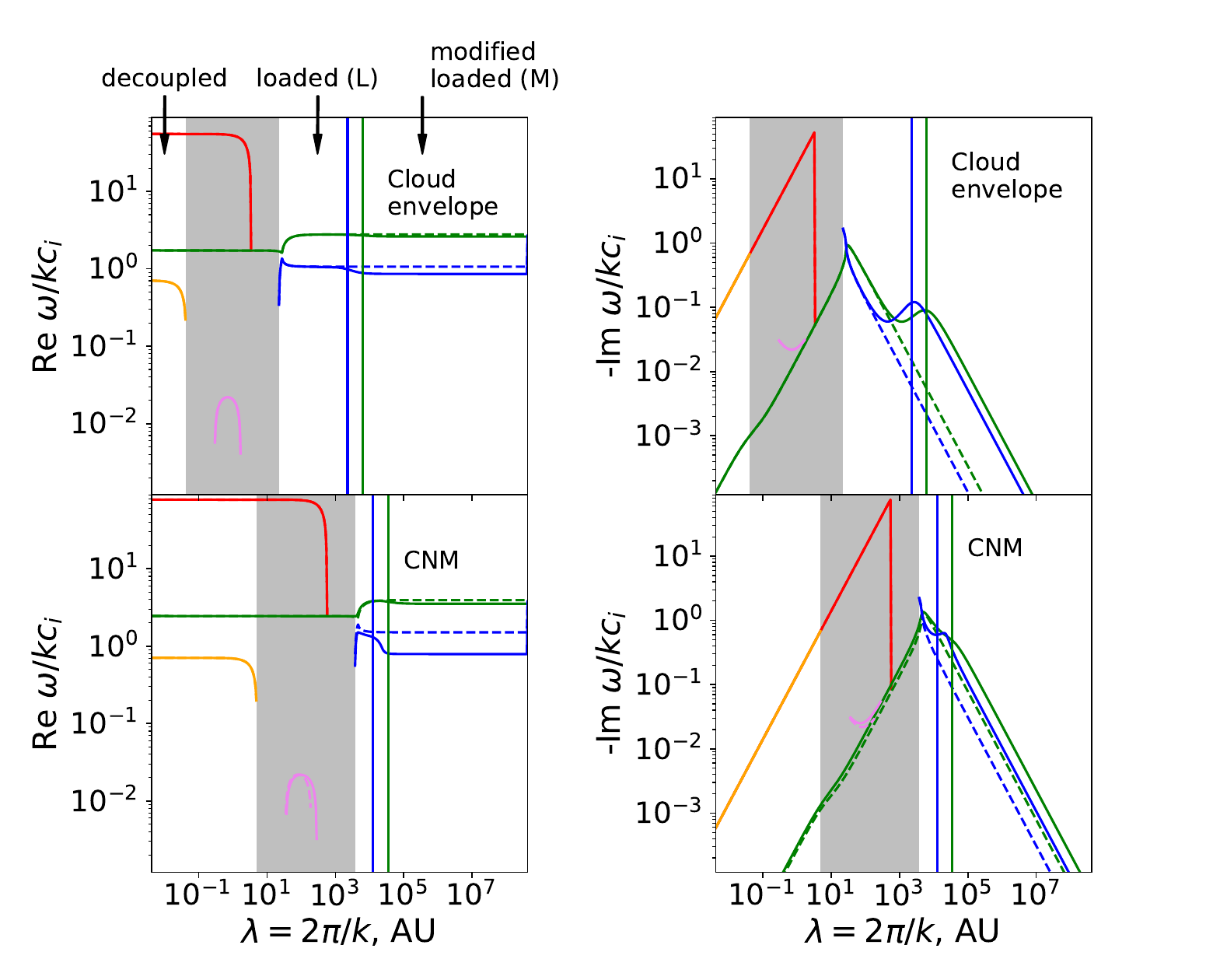}
\caption{Illustration of the dispersion relation of magnetosonic modes derived from Equations~\eqref{eq20}--\eqref{eq21} for two phases of ISM. The solid curves show the real and imaginary parts of the dispersion relation plotted as a function of the
wavelength $\lambda$.  The dashed curves show the result in the limit $\tau_{\rm c} \rightarrow \infty$ studied in
\citet{Soler13}.  Here, we have assumed an angle $\theta$ of 45$^\circ$ between ${\bf k}$ and ${\bf B}$, and that the loaded
Alfv{\'e}n speed $c_{\rm LA} = 1.4 c_n$ (corresponding to plasma $\beta = 0.6$). The shaded gray region indicates the
decoupling gap -- the wavelength range over which the decoupling of ions and neutrals occurs. The blue and green vertical
lines show the wavelengths where Re~$\omega = \tau_{\rm c}^{-1}$, near where the thermal damping peaks for the corresponding mode.
The decoupled, loaded (L), and modified loaded (M) regimes are indicated above the top left panel (see Section
\ref{sect:effect} for a detailed discussion).} \label{fourPanel}
\end{figure*}

In this section, we numerically calculate the dispersion relation on the basis of Equations \eqref{eq20}--\eqref{eq21}, with
and without thermal damping, and analyze important features introduced by the damping. The results are illustrated for
characteristic astrophysical environments.  In this paper, we have introduced a number of parameters, including in particular speeds of wave modes for three different propagation regimes discussed in this section (see Figure \ref{fourPanel}).  The most important of these parameters are listed in Table~\ref{tab:table1}.  

\begin{table}[h!]
  \begin{center}
    \caption{{\bf Notation for the key parameters}}
    \label{tab:table1}
  \begin{tabular}{c c c }
        \hline
        Parameter & {Meaning} & {Equation} \\
        \hline
     $\tau_{\rm c}$ & gas cooling time & \eqref{psiAndAlpha} \\
     $\nu_{in}$ & ion-neutral collision frequency & \eqref{eq:nus} \\
     $\nu_{ni}$ & neutral-ion collision frequency & \eqref{eq:nus} \\
     $c_{\rm A}$ & Alfv{\'e}n speed & \eqref{eq:cA} \\
     $c_{\rm LA}$ & loaded Alfv{\'e}n speed & \eqref{eq:cLA} \\
     $c_i$ & ion sound speed & \eqref{eq:ci} \\
     $c_n$ & neutral sound speed & \eqref{eq:cn} \\
     $c_{{\rm M}n}$ & modified sound speed & \eqref{eq:cmn} \\
     $c_{\rm f, s}$ & fast/slow speed & \eqref{eq:fastMode} \\
     $c_{\rm Lf,s}$ & loaded fast/slow speed & \eqref{eq:fastMode} with \\ ~ & ~ &$c_{\rm A} \rightarrow c_{\rm LA}$,\\ ~ & ~ &$c_i \rightarrow c_n$ \\
     $c_{\rm Mf, s}$ & modified loaded fast/slow speed & \eqref{eq:fastMode} with \\ ~ & ~ & $c_{\rm A} \rightarrow c_{\rm LA}$,\\ ~ & ~ &$c_i \rightarrow c_{{\rm M}n}$\\
     $\chi$ & ratio $(C_{{\rm M}n}/c_n)^2$  & \eqref{eq:chi} \& \eqref{eq:cmn} \\
     $\gamma$ & adiabatic index & $C_p/C_v$ \\
     $\beta$ & thermal to magnetic pressure ratio & \eqref{eq:beta} \\
     $\beta_{\rm M}$ & $\beta$ in the modified loaded regime & \eqref{eq:betaStar}\\
     $\mathcal{M}_n$ & sonic Mach number & $v_0/c_{n}$ \\
     $\mathcal{M}_{\rm LA}$ & (loaded) Alfv{\'e}nic Mach number & $v_0/c_{\rm LA}$\\
      \hline
    \end{tabular}
  \end{center}

\end{table}

\subsection{Astrophysical Environments}
\label{physEnv}

In our analysis, we study the effect in detail in two phases of the ISM: the cold neutral medium (CNM), and the
envelope of a molecular cloud. We discuss the expected damping in other environments in Section \ref{sect:magnitude}.

\subsubsection{CNM}
\label{sect:CNM}
The CNM is made up of atomic hydrogen, for which we assume $n_n= 20$~cm$^{-3}$. The ion is C$^+$, and the ionization fraction is $1.6 \times 10^{-4}$.  The presence of helium or other elements is ignored.  Heating is dominated by the photoelectric effect on dust, and given by $\Lambda_{\rm h} = 1.4 \times 10^{-26}(n_n/{\rm cm}^{-3})$~erg~cm$^{-3}$~s$^{-1}$ \citep{Draine11}.  The cooling is from fine-structure lines of C$^+$, and we use the optically thin cooling prescription considering collisions with H atoms and electrons from \citet{Gong17}.
 This gives $T_0=60$~K, $\tau_{\rm c}=6 \times 10^{11}$~s,
and $\chi=0.22$.

\subsubsection{Cloud Envelope}
\label{sect:subsubCloudEnv}
The neutral constituent is assumed to be molecular hydrogen, the ion is C$^+$, and the ionization fraction is 0.032\% (implying, as in the CNM, a C/H ratio of 0.016\%).  \citet{Goldsmith01} suggests that $\Lambda_{\rm c} = 4.4 \times10^{-27}(T/{\rm K})^{2.4}$~erg~cm$^{-3}$~s$^{-1}$ for $n_n=10^3$~cm$^{-3}$. We assume heating to be from a combination of CR and photoelectric heating, with a rate given by $\Lambda_{\rm h} = 1.0 \times 10^{-26}(n_n/{\rm cm}^{-3})$~erg~cm$^{-3}$~s$^{-1}$. Setting $n_n=10^3$~cm$^{-3}$, we obtain $T_0=25$~K.  At this temperature, the logarithmic slope of the cooling function with density, estimated from the data in Table 2 from \citet{Goldsmith01}, is approximately 1.1.  Applying Equations \eqref{psiAndAlpha} and \eqref{eq:chi}, we find  $\tau_{\rm c}=2.2 \times 10^{11}$~s, and $\chi=0.58$.

\subsection{Effect of Thermal Damping}
\label{sect:effect}

Figure \ref{fourPanel} illustrates the dispersion relation for magnetosonic waves, showing a comparison of the cases with and
without the thermal damping effect. For all panels, we set an angle $\theta$ between ${\bf k}$ and ${\bf B}_0$ of 45$^\circ$
and assume that what we call the {\it loaded} Alfv{\'e}n speed, given by
\begin{equation}
c_{\rm LA} = \frac{B_0}{\sqrt{4\pi \rho}} \equiv c_{\rm A} \sqrt{\frac{\rho_i}{\rho}}\,.
\label{eq:cLA}
\end{equation}
is equal to 1.4 times the neutral sound speed.  This is the speed of Alfv{\'e}n waves in the regime where collisions are so frequent that the neutrals are dragged along by
the ions, i.e., everything to the right of the gray region in Figure \ref{fourPanel}.  The coefficient 1.4 was derived
from the respective magnetic and thermal energies indicated in Table 1.5 of \citet{Draine11}.  This corresponds to a plasma
$\beta$ of 0.6, where $\beta$ is the ratio of thermal to magnetic pressure:
\begin{equation}
    \beta = \frac{8\pi n k_{\rm B} T_0}{B_0^2} \equiv \frac{2c_n^2}{\gamma c_{\rm LA}^2}\,.
    \label{eq:beta}
\end{equation}
The dispersion relation is plotted as a function of wavelength $\lambda = 2\pi/k$. We have only plotted propagating modes
(those with a non-zero real part). In each case, there is a corresponding mode (not shown) with a negative real part of
equal magnitude.

The solid lines in Figure~\ref{fourPanel} show the dispersion relation of magnetosonic modes derived from Equations \eqref{eq20}--\eqref{eq21}. The
dashed lines depict the results in the adiabatic limit $\tau_{\rm c} \rightarrow \infty$ [where $f(\omega) \rightarrow 1$ in
Equation \eqref{eq12}]. In all illustrated cases, there is one mode, shown in green, which is {\it continuously present} at
all $\lambda$. For the chosen parameters, this is the loaded fast mode in the long-wavelength regime, which becomes the
neutral acoustic mode in the short-wavelength (decoupled) regime (see Section~\ref{sec:cont} and Appendix~\ref{App_gap} for
general analysis). Also plotted are the loaded slow mode (blue), the decoupled fast and slow modes (red and orange), as well
as a low-frequency mode (violet), present at intermediate wavelengths.

The speeds of the decoupled fast and slow modes, $c_{\rm f,s}$, can directly obtained from Equations~\eqref{eq20} and
\eqref{eq21} by neglecting the frictional coupling terms. This yields the classical expression for magnetosonic modes \citep{Landau60}
\begin{equation}
 c_{\rm f, s}^2 =  \frac{1}{2} \left[c_{\rm A}^2 + c_i^2 \pm \sqrt{(c_{\rm A}^2 + c_i^2)^2 - 4 c_{\rm A}^2 c_i^2\cos^2{\theta}}\:\right].
 \label{eq:fastMode}
\end{equation}
In the loaded regime, where ions and neutrals are moving together, the speeds $c_{\rm Lf,s}$ of loaded fast and slow modes
are given by Equation~\eqref{eq:fastMode} where $c_{\rm A}$ is replaced by $c_{\rm LA}$, and $c_i$ is replaced by $c_n$. The range
of wavelengths over which ion-neutral decoupling occurs -- referred below as the {\it decoupling gap} -- is indicated in
each panel of Figure~\ref{fourPanel} by the shaded gray region. As shown in Appendix \ref{App_gap}, the right edge of the
decoupling gap is approximately described by the condition $kc_{\rm LA} \approx 2\nu_{ni}\cos\theta$ for $c_{\rm LA}\ll
c_n$ (where $c_{\rm Lf}\approx c_n$ and $c_{\rm Ls}\approx c_{\rm LA}\cos\theta$), and by $kc_{\rm LA} \approx 2\nu_{ni}$ in the opposite limit (where $c_{\rm Lf} \approx c_{\rm LA}$ and $c_{\rm Ls} \approx c_n \cos{\theta}$).  In Figure
\ref{fourPanel}, the edges of the decoupling gap were determined numerically. In Appendix~\ref{App_gap} we present a detailed
analysis of wave modes in the loaded regime and their connection to the decoupling gap.

The blue and green vertical lines on the right side from the decoupling gap correspond to the condition Re~$\omega
=\tau_{\rm c}^{-1}$, where $\omega$ is the frequency represented by the curve of the same color.  These lines approximately match
the peak of the damping associated with radiative cooling, where the real parts of the loaded fast and slow modes change to
their {\it modified loaded} values.  These are given by Equation \eqref{eq:fastMode}, with $c_{\rm A}$ replaced by $c_{\rm LA}$ (as in the loaded regime), while
$c_i$ is now replaced by
\begin{equation}
 c_{{\rm M}n} = c_n \sqrt{\chi}\,,
 \label{eq:cmn}
 \end{equation}
as follows from Equations \eqref{eq12} and \eqref{eq:fdef} in the low-frequency limit.

We see that thermal damping is only important in the modified loaded regime, at $\lambda$ much larger than the scale of ion-neutral
decoupling.  This justifies our initial assumption that ions and neutrals are thermally coupled.  For the chosen parameters,
thermal damping has a rather moderate effect on the real part of both modes, but a dramatic effect -- ranging from a factor
of several in the CNM to several hundred in the case of the cloud envelope -- on the imaginary part.

\subsection{Thermal Damping versus Propagation Angle} \label{sect:Angle}

Figure \ref{fourPanel} suggests that thermal damping can increase the imaginary part of wave modes in the modified loaded regime by
over two orders of magnitude. For that reason, in this section we focus on the damping rate and explore how that depends on
propagation angle. In Appendix~\ref{app:modifiedDamping} we show that the imaginary part of the modified loaded fast and slow modes
is approximately given by
\begin{equation}
    -{\rm Im}~\omega = \frac{\chi^{-1} - 1}4 \,\tau_{\rm c} c_{{\rm M}n}^2 \mathscr{F}_{\rm f, s}(\beta_{\rm M}, \theta)k^2,
    \label{eq:dampAppend}
\end{equation}
where $\mathscr{F}_{\rm f, s}$ is given by Equation~\eqref{eq:fancyF} and
\begin{equation}
\beta_{\rm M} = \frac{c_{{\rm M}n}^2}{c_{\rm LA}^2} \equiv \frac{\chi \gamma}2 \beta,
\label{eq:betaStar}
\end{equation}
is the plasma $\beta$ in the modified regime.

\begin{figure}
\centering
\includegraphics[width=1.0\columnwidth]{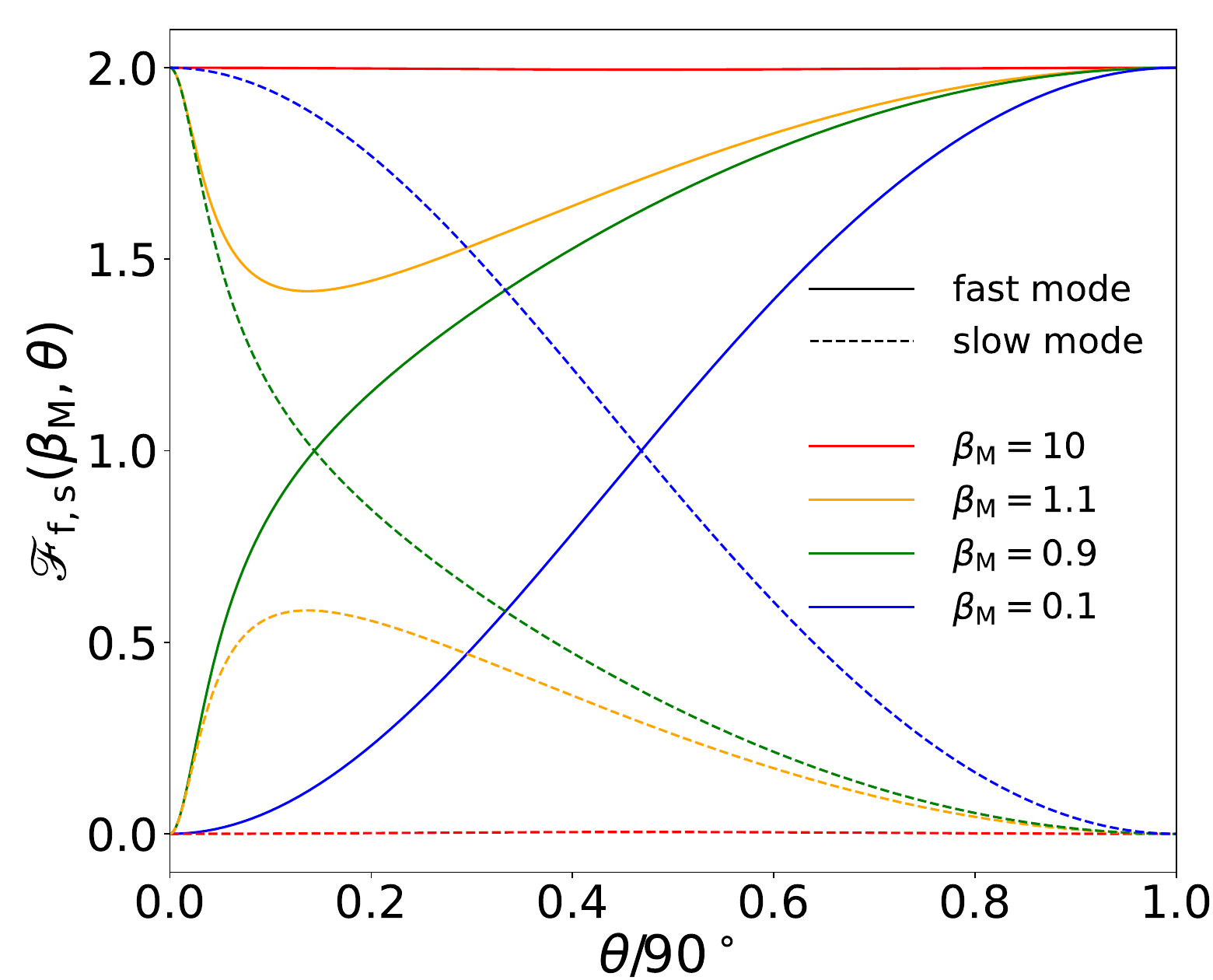}
\caption{Prefactor $\mathscr{F}_{\rm f,s}$ of the thermal damping rate, given by Equation~\eqref{eq:dampAppend}.
Curves are potted as a function of $\theta$ for different values of $\beta_{\rm M} = \chi\gamma\beta/2$.  The solid lines
correspond to the modified fast loaded mode $\mathscr{F}_{\rm f}$, and the dashed lines to the modified slow loaded mode, $\mathscr{F}_{\rm s} =
2 - \mathscr{F}_{\rm f}$.}
\label{appendixFigure}
\end{figure}

The strength of thermal damping is related to the degree to which the mode results in compression of the gas. Therefore, we
expect the imaginary part of the dispersion relation in the modified loaded regime to be a function of $({\bf k} \cdot {\bf v})^2$,
where ${\bf v}$ is the velocity eigenvector.  The components $v_\perp$ and $v_\parallel$ of the eigenvector, perpendicular
and parallel to ${\bf k}$, are related via \citep[e.g.,][]{Landau60}
\begin{equation}
    \frac{v_\perp}{v_\parallel}=\frac{\cos{2\theta}-\beta_{\rm M} \pm\sqrt{1+ \beta_{\rm M}^2-2\beta_{\rm M} \cos{2\theta}}}{\sin{2\theta}}\,,
    \label{eq:velocityDirection}
\end{equation}
for the fast ($+$) and slow ($-$) mode. Using Equations~\eqref{eq:velocityDirection} and \eqref{eq:fancyF}, we find that
\begin{equation}
    \mathscr{F} = 2 (\hat{\bf k} \cdot \hat {\bf v})^2.
\end{equation}
This dependence allows us to easily understand the behavior of $\mathscr{F}_{\rm f, s}$, plotted in
Figure~\ref{appendixFigure} as a function of $\theta$ for different values of $\beta_{\rm M}$. For large $\beta_{\rm M}$ (i.e., weak
magnetization) the fast mode is approximately the (modified) neutral sound mode, with ${\bf v}$ almost parallel to ${\bf
k}$, while ${\bf v}$ of the slow mode (with $c_{\rm Ms}\approx c_{\rm LA}\cos\theta$) is almost perpendicular to ${\bf k}$,
as follows from Equation~\eqref{eq:velocityDirection}. Therefore, the damping of the fast mode is strong and almost
independent of $\theta$ ($\mathscr{F}_{\rm f}\approx2$), while the damping of the slow mode is weak. On the contrary, for small
$\beta_{\rm M}$ (strong magnetization) we have $c_{\rm Mf}\approx c_{\rm LA}$ with ${\bf v}$ almost perpendicular to ${\bf B}_0$,
and $c_{\rm Ms}\approx c_{{\rm M}n}\cos\theta$ with ${\bf v}$ parallel to ${\bf B}_0$. In this case, we have $\mathscr{F}_{\rm
f}\approx2\sin^2\theta$ and $\mathscr{F}_{\rm s}\approx2\cos^2\theta$. We note that the damping of the fast and slow modes
is symmetrically opposite, generally following $\mathscr{F}_{\rm f}+\mathscr{F}_{\rm s}=2$.

\subsection{Thermal Damping of the Continuous Mode}
\label{sec:cont}

\begin{figure}[htp]
\centering
\includegraphics[width=1.0\columnwidth]{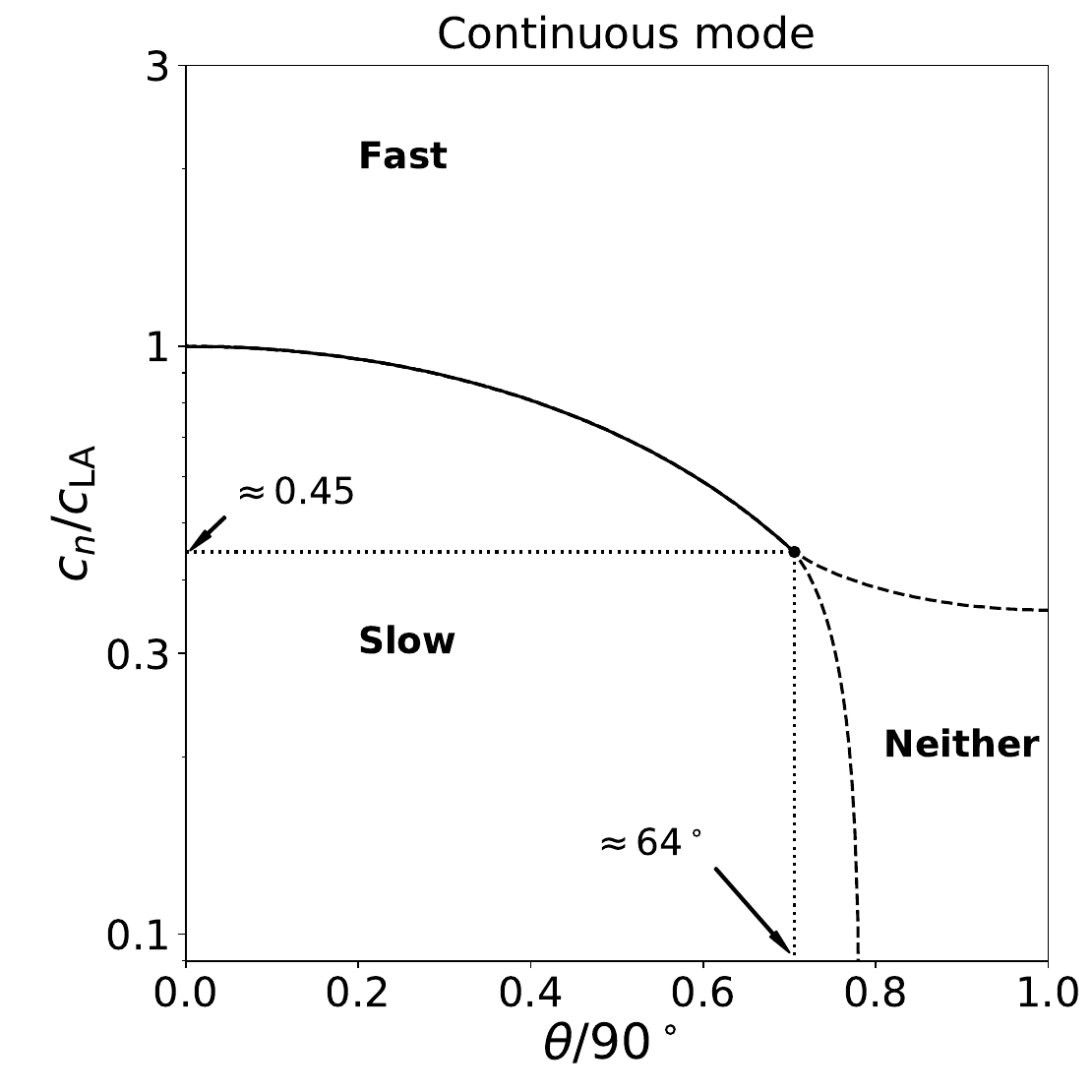}
\caption{The parameter space of $c_n/c_{\rm LA}\equiv\sqrt{\beta_{\rm M}/\chi}$ and $\theta$, showing where either the
fast mode, the slow mode, or neither mode are continuous with the neutral sound mode in the decoupled regime (see the left
panel of Figure~\ref{fourPanel}). The solid line is the analytical condition $c_n/c_{\rm LA}=\cos\theta$,
Equation~\eqref{switch}, the dotted lines are computed numerically.} \label{continuityImage}
\end{figure}

Let us determine a condition for the existence of the {\it continuous mode}. In all environments considered in this paper,
$\rho_i \ll \rho_n$.  In this limit, if there is a continuous mode, it is the neutral sound mode in the decoupled regime and
throughout the decoupling gap. In Appendix~\ref{App_gap} we derive that, depending on parameters, it can switch either to
the loaded fast mode (as illustrated in Figure~\ref{fourPanel}) or to the loaded slow mode at the right edge of the
decoupling gap. Figure~\ref{continuityImage} shows that the continuity is determined by two parameters, $c_n/c_{\rm LA}$ and
$\theta$. For $\theta\lesssim64^\circ$, the neutral sound branch switches between the loaded fast and slow branches at the
solid line, described by Equation~\eqref{switch}. However, the neutral sound mode disappears completely within the
decoupling gap for sufficiently small $c_n/c_{\rm LA}$ and large $\theta$, i.e., for the parameters bound in the right
bottom corner of Figure~\ref{continuityImage} by the dashed lines, no mode is continuous for all $k$.

Now we can evaluate the damping rate of the continuous mode in the modified loaded regime. From Figures~\ref{appendixFigure} and
\ref{continuityImage} we conclude that for large $c_n/c_{\rm LA}\equiv\sqrt{\beta_{\rm M}/\chi}$, the continuous mode in the
modified loaded regime is represented by the fast mode, whose damping is strong and almost independent of $\theta$. For small
$c_n/c_{\rm LA}$, the continuous mode (for $\theta$ where it exists) is the slow mode, which is strongly damped too
(although the damping decreases with $\theta$). In the range of $0.45\lesssim c_n/c_{\rm LA}<1$, the continuous mode
switches from the slow to the fast mode as $\theta$ increases. The damping is then described by the dashed green or blue
lines in Figure~\ref{appendixFigure} at smaller $\theta$, switching to the respective solid lines at larger $\theta$.

Something qualitatively different occurs when $1<c_n/c_{\rm LA}<1/\sqrt{\chi}$. In this range, the continuous mode is the
fast mode, but we see in Figure \ref{appendixFigure} (where the corresponding range is $\chi < \beta_{\rm M}<1$) that the fast
mode is then weakly damped at small $\theta$.  This implies that the resulting turbulent spectrum is expected to be strongly
anisotropic for this range of $c_n/c_{\rm LA}$.

We conclude that thermal damping of the continuous mode -- as long as it exists -- is generally strong. The damping rate can
vary with $\theta$, depending on the value of $c_n/c_{\rm LA}$, but this variation is rather insignificant outside the
relatively narrow range of $1<c_n/c_{\rm LA}<1/\sqrt{\chi}$. In the next section, where we analyze the impact of thermal
damping on the turbulent cascade, the damping rate is therefore assumed to be independent of $\theta$.

\section{Characteristics of Thermal Damping in Different Environments}
\label{sect:magnitude}

In assessing the importance of thermal damping, we consider two criteria.  First is the peak value of $\tilde\omega_I \equiv {\rm |Im \, \omega|}/{\rm Re \, \omega}$ due to thermal damping, providing a measure of the overall strength of the effect.  Second, we consider the length scale at which the effect is strongest, and compare this with the scale corresponding to the right edge of the decoupling gap.  Magnetosonic modes are generally affected by both thermal damping and ion neutral friction, so it is of interest to ask which mechanism will damp the turbulent cascade at larger scales.  To address this question, in this section we also derive the ratio of thermal damping rate to damping rate due to ion-neutral friction in the long-wavelength limit.  

\begin{figure}
\centering
\includegraphics[width=1.0\columnwidth]{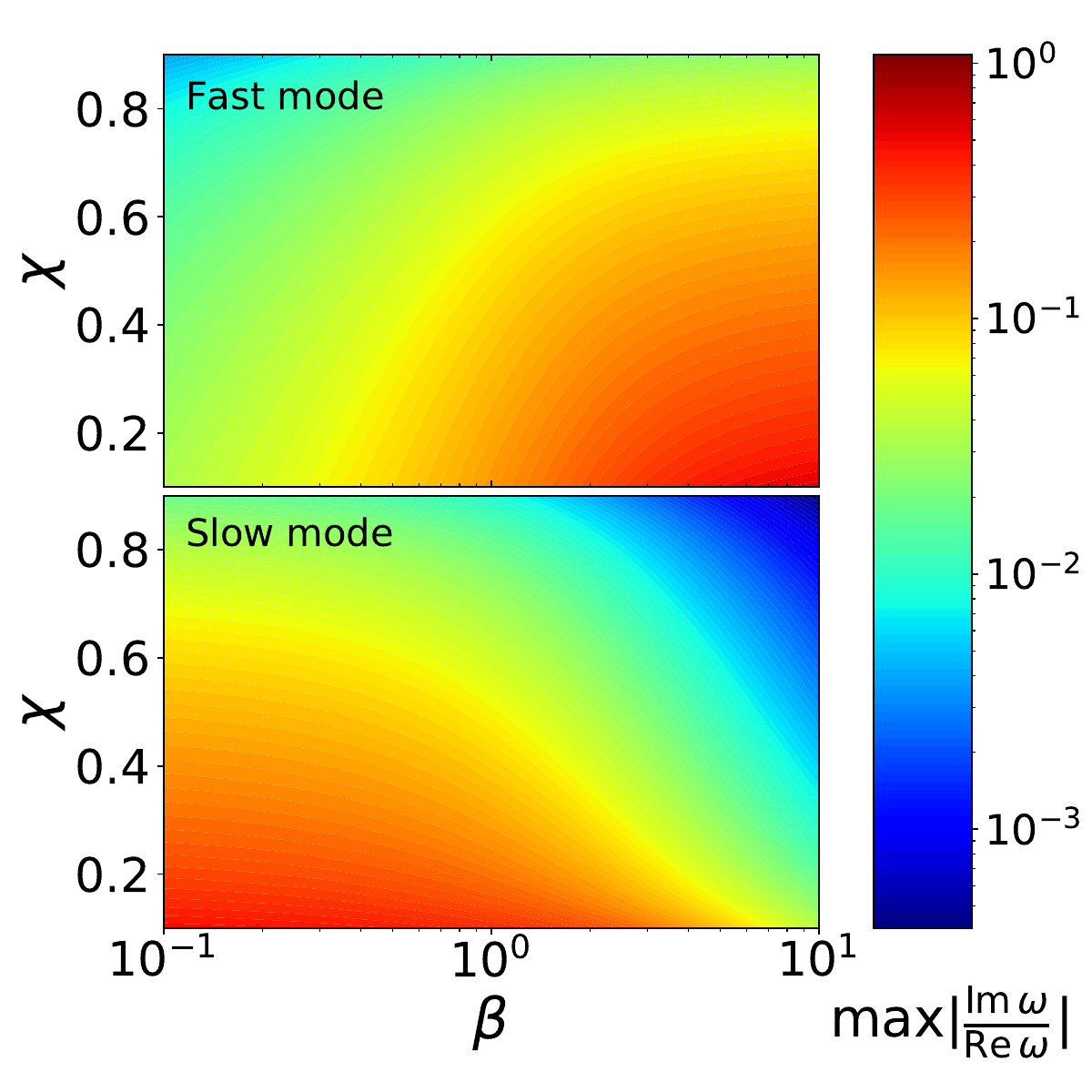}
\caption{Peak value of $\tilde \omega_I = {\rm |Im \,\omega|}/{\rm Re \, \omega}$ due to thermal damping, plotted in the plane of $\beta$ and $\chi$.  As expected from the considerations discussed in Sections~\ref{sect:Angle} and \ref{sec:cont}, the peak of $\tilde \omega_I$ is higher for the continuous mode, which is represented by the fast mode if $\beta \gg1$, and by the slow mode if $\beta \ll 1$.  In both cases, the effect is maximized for small $\chi$. }
\label{fig:peakValue}
\end{figure}

Equation \eqref{eq:disp5} describes the dispersion relation in the absence of ion-neutral friction.  We numerically solved Equation \eqref{eq:disp5} to determine the peak value of $\tilde \omega_I$ as a function of $\beta$ and $\chi$.  This is shown for both fast and slow modes in Figure \ref{fig:peakValue}.  We note that the peak value of $\tilde \omega_I$ does not depend on $\tau_{\rm c}$.  To see this, define $\omega(k, \tau_{\rm c})$ as a solution of Equation \eqref{eq:disp5} for given parameters $\beta$ and $\chi$.  We note that $\omega(k/x, x \tau_{\rm c}) = x^{-1} \omega(k, \tau_{\rm c}$) for an arbitrary constant $x$.  This means that the ratio $\tilde \omega_I$ satisfies $\tilde \omega_I(k/x, x \tau_{\rm c}) = \tilde \omega_I(k, \tau_{\rm c})$.  Therefore, by changing $\tau_{\rm c}$, the location of the peak value of $\tilde \omega_I$ is shifted to different $k$ such that $k \tau_{\rm c}$ remains constant, but the peak value of $\tilde \omega_I$ is unchanged. 

As discussed in Section \ref{sect:Angle}, $\tilde \omega_I$ depends on the mode compressibility, and this behavior is reflected in Figure \ref{fig:peakValue}.  Modes which are mostly compressive have higher peak values of $\tilde \omega_I$.  These are the fast modes in the limit of high $\beta$ and the slow modes in the limit of low $\beta$ (see Figure \ref{appendixFigure}).  Figure \ref{fig:peakValue} also shows that the effect is maximized for small values of $\chi$.  This is consistent with Equation \eqref{eq:dampAppend}, which shows that the damping rate of fast and slow modes is proportional to the deviation of $\chi^{-1}$ from unity. 

The question of the wavelength $\lambda_{\rm peak}$ at which damping peaks relative to the right edge of the decoupling gap, $\lambda_{\rm dec}$, is more difficult to address in general.  The physics which determines the cooling is complicated and varies throughout the different ISM phases.  In addition to the two environments analyzed above in detail (the CNM and cloud envelope), in this section we also briefly discuss the degree of thermal damping in the warm neutral medium (WNM), and the inner and outer parts of a prestellar core.

\subsection{Different additional environments}

In the WNM, the ion is H$^+$ and the neutral component is atomic hydrogen.  The heating in the WNM is dominated by the photoelectric effect on dust, and we adopt a constant heating rate of $\Lambda_{\rm h} = 1.4 \times 10^{-26}(n_n/{\rm cm}^{-3})$~erg~cm$^{-3}$~s$^{-1}$ from \citet{Draine11}.  The cooling function $\Lambda_{\rm c}$ is estimated from  Figure~30.1 of \citet{Draine11}, where an ionization fraction of 1.7\% is assumed, and we adopt this value, too.  We set the gas number density $n_n
= 0.4$~cm$^{-3}$, and calculate an equilibrium gas temperature $T_0=8.7 \times 10^3$~K. This yields a gas cooling time $\tau_{\rm c}=5 \times 10^{13}$~s, and $\chi=0.37$.  

The right edge of the decoupling gap is approximated by $k c_{\rm LA} \approx 2\nu_{ni}$ (see the end of Appendix \ref{App_gap}).  This gives the decoupling scale $\lambda_{\rm dec}= \pi c_{\rm LA}/\nu_{ni}$, which corresponds to $2\times 10^4$ AU in the WNM. The peak of the thermal damping is estimated from the criterion $\omega \approx \tau_{\rm c}^{-1}$. 
Since the fast mode is the one which is continuous over all $\lambda$ for our fiducial values of $\beta=0.6$ and $\theta=45^\circ$,
we obtain $\lambda_{\rm peak} \approx 2 \pi \tau_{\rm c} c_{\rm Lf}$, which 
corresponds to a scale of $4 \times 10^7$ AU in the WNM.

The continuous mode in the WNM is also affected by collisionless damping, because the thermal speed of ions in this environment is about $c_n$.  Therefore, were the medium fully ionized, the rate of collisionless damping due to ions would be of order $\omega$ for most values of $\theta$ \citep{Ginzburg70}.  However, since the wave energy is proportional to the mass of neutrals, but only ions contribute to the damping, the resulting damping rate (in units of $\omega$) is bounded by the ionization fraction, which we assume to be 1.7\%. 

Let us also consider different regions of a prestellar core. Adjacent to the cloud envelope is what we call an {\it outer core} -- a region where the interstellar UV radiation is already attenuated, so that the ionization fraction is controlled by CRs, but the gas density is still sufficiently low to ensure the cooling dominated by gas species. A denser region, where the cooling is due to dust, is referred to as an {\it inner core}.

For the outer core, we set $n_n = 10^4$ cm$^{-3}$. The heating is assumed to be dominated by CRs, with $\Lambda_{\rm h} = 1.0 \times
10^{-27}\, (n_n/{\rm cm}^{-3})$~erg~cm$^{-3}$~s$^{-1}$ (i.e. an order of magnitude smaller than in Section \ref{sect:subsubCloudEnv} for a given value of $n_n$). Again, we use the cooling function taken from Table 2 of \citet{Goldsmith01}, appropriate for a density of $10^4$~cm$^{-3}$, which yields $T_0=12.4$~K, $\tau_c = 1.0 \times 10^{12}$ s, and $\chi = 0.67$. HCO$^+$ is assumed to be the principal ion with fractional abundance of $10^{-7}$ relative to H$_2$ \citep{Williams98}. We obtain $\lambda_{\rm peak} \approx 2 \times 10^4$~AU and $\lambda_{\rm dec} \approx 4 \times 10^3$~AU. As before, $\lambda_{\rm peak}$ and $\lambda_{\rm dec}$ are evaluated for the fast mode, assuming $\beta = 0.6$ and $\theta = \pi/4$.

The inner core corresponds to gas densities above $10^5$ cm$^{-3}$, where the equilibrium thermodynamics is determined by CR heating and cooling due to collisions with dust grains. As discussed in \citet{Ivlev19}, the rate of cooling generally depends on the gas density and dust size distribution. In particular, there is a critical grain radius $A$, given by 
\begin{equation}
    A = \frac{1}{\sqrt{18\pi}} \frac{n_nv_nk_{\rm B}}{q_{\rm abs}\sigma T_{d0}^5}\,,
\end{equation}
where $v_n = \sqrt{k_{\rm B}T/m_n}$ is the scale of gas thermal velocity for the mass $m_n = 2m_{\rm H}$, $q_{\rm abs}$ is a material-dependent factor taken to be 0.13 K$^{-2}$ cm$^{-1}$, $\sigma$ is the Stefan-Boltzmann constant, and $T_{\rm d0}$ is the equilibrium dust temperature, set by the local radiation field in the absence of coupling to the gas.

\begin{table*}
  \begin{center}
    \caption{{\bf Characteristics of thermal damping}$^{\rm a}$\\
    \vspace{.3 cm}
    \hspace{1 cm} Environment  \hspace{4cm} $c_{\rm LA}/c_{n} = 1.4~(\beta=0.6)$ \hspace{2.5cm} $c_{\rm LA}/c_{n} = 0.5~(\beta=4.8)$}
    \label{tab:table2}
    \vspace{-.2 cm}
  \begin{tabular}{c c c c | c c c | c c c }
        \hline
         &  $T$, K & $n_n$, cm$^{-3}$ & $n_i/n_n$ & $\lambda_{\rm dec}$, AU & $\lambda_{\rm peak}$, AU & $\mathscr{R}$ & $\lambda_{\rm dec}$, AU & $\lambda_{\rm peak}$, AU & $\mathscr{R}$ \\
        \hline
     WNM & 8700 & 0.4 & 0.017 & $2 \times 10^4$ & $4 \times 10^7$ & 140 & 9000 & $2 \times 10^7$ & 2400 \\
     CNM & 60 & 20 & $1.6 \times 10^{-4}$ & 3000 & $4 \times 10^4$ & 1.2 & 1200 & $2 \times 10^4$ & 28 \\
     Cloud envelope & 25 & 1000 & $3.2 \times 10^{-4}$ & 20 & $6 \times 10^3$ & 23 & 7 & 4000  & 300 \\
     Outer core & 12.4 & $10^4$ & $10^{-7}$& $4 \times 10^3$ & $2 \times 10^4$ & 0.3 & 1500 & $1.3 \times 10^4$ & 3 \\
     Inner core& 6 & $\gtrsim 10^5$ & Eq. \eqref{ionizationEquation} & $1300/n_6^{0.44}$ & $160\,a_{0.1}/n_6$ & $0.01
     \,a_{0.1}/n_6^{0.56}$ & $450/n_6^{0.44}$ & $100
     \,a_{0.1}/n_6$ & $0.1
     \,a_{0.1}/n_6^{0.56}$ \\
     ($n_6/a_{0.1} \ll 2$) &~&~&~ & ~ & ~& ~ & ~ & ~& ~\\
     Inner core & 6 & $\gtrsim 10^5$ & Eq. \eqref{ionizationEquation} & $1300/n_6^{0.44}$ & 80 & $0.004\,n_6^{0.44}$ & $450/n_6^{0.44}$ & 50 & $0.06\,n_6^{0.44}$ \\
     ($n_6/a_{0.1} \gg 2$) &~&~&~  & ~ & ~ & ~ & ~ & ~ & ~ \\
        \hline
    \end{tabular}
  \end{center}
 $^{\rm a}$Wavelengths of the decoupling and damping peaks are calculated as $\lambda_{\rm dec}= \pi c_{\rm LA}/\nu_{ni}$ and $\lambda_{\rm peak}= 2\pi \tau_{\rm c} c_{\rm Lf}$, the damping rate ratio $\mathscr{R}$ (for the continuous mode) is given by Equation~\eqref{eq:scriptR}; $n_6$ is the gas density $n_n$ in units of $10^6$ cm$^{-3}$, $a_{0.1}$ is the effective grain radius $a_{\rm eff}$ in units of 0.1 microns.

\vspace{.2cm}

\end{table*}

If the effective grain radius $a_{\rm eff}$ --- see Equation 16 of \citet{Ivlev19} --- is much larger than $A$, then the grain temperature is independent of the gas temperature.  This regime occurs when $n_6/a_{0.1} \ll 2$, where $n_6$ is $n_n$ in units of $10^6$ cm$^{-3}$ and $a_{0.1}$ is $a_{\rm eff}$ in units of 0.1 microns. In this regime, we estimate the cooling from Equation 18 of \citet{Ivlev19}.  Using Equations \eqref{psiAndAlpha} and \eqref{eq:dustCooling}, assuming $T - T_{d0} \ll T_{d0}$, and keeping in mind that $C_v = 3k_{\rm B}/(4m_{\rm H})$ for molecular hydrogen, this leads to a cooling time of $\tau_{\rm c} \approx \sqrt{\pi/8} \: a_{\rm eff}\rho_d/(f_d v_n \rho_n) = 1.2 \times 10^{10} n_6^{-1} a_{0.1}$~s. In this expression, $f_d$ is the dust to gas mass ratio, $\rho_n$ the gas mass density, and $\rho_d$ the {\it material} density of the dust grains.  In the numerical estimate, we have taken $f_d = 0.02$ and $\rho_d = 2$ g cm$^{-3}$.  We have chosen such values of $f_d$ and $\rho_d$ to account for the fact that the grains are likely covered with icy mantles which increase their mass fraction and lower their mean density.  Molecular cloud cores show a wide range of $\beta$ \citep{Crutcher10}. For $\beta = 0.6$, as in Figure \ref{fourPanel}, the corresponding peak damping scale is $\lambda_{\rm peak} = 2 \pi \tau_{\rm c} c_{\rm LF}\approx 160 n_6^{-1} a_{0.1}$ AU. Assuming $T - T_{d0} \ll T_{d0}$, Equation \eqref{eq:chi} yields $\chi \approx \gamma^{-1} = 3/5$.  

In the opposite limit, where $n_6/a_{0.1} \gg 2$, the cooling can be obtained from Equation (19) of \citet{Ivlev19}:
\begin{equation}
    \Lambda_{\rm c}(\rho_n, T) = \frac{3 f_d \rho_n q_{\rm abs} \sigma }{\rho_d} \left(T^6 - T_{\rm d0}^6\right),
    \label{eq:dustCooling}
\end{equation}
while $\Lambda_{\rm h}(\rho_n, T) \propto \rho_n$.  From this, we calculate $\tau_{\rm c} = k_{\rm B} \rho_d/(24 f_d m_{\rm H} q_{\rm abs}\sigma T^5) = 6.0 \times 10^{9}$ s, assuming 
$T_{\rm d0} = 6$~K. Setting, as before, $\beta = 0.6$, this results in $\lambda_{\rm peak} = 2 \pi \tau_{\rm c} c_{\rm Lf} \approx 80$ AU (independent of $n_n$). Equation \eqref{eq:chi} yields $\chi \approx \gamma^{-1} = 3/5$ in this limit, too.

To estimate the right edge of the decoupling gap in the inner core, we must have an estimate for the density and ionization fraction.  We use the relation given for the ionization fraction as a function of density in \citet{Caselli02} (their model 3), 
\begin{equation}
    n_i/n_n = 2.3\times 10^{-9} n_6^{-0.56},
    \label{ionizationEquation}
\end{equation}
and assume the dominant ion to be HCO$^+$.  The criterion $k c_{\rm LA} = 2 \nu_{ni}$ yields the scale of the right edge of the decoupling gap in terms of density: $\lambda_{\rm dec} = 1.3 \times 10^3$ $n_6^{-0.44}$ AU.  

We conclude that, although there are significant sources of uncertainty, the peak of the thermal damping in the inner core occurs at scales substantially smaller than the right edge of the decoupling gap. 
 
\subsection{Thermal Damping versus Ion-neutral Friction}

One quantity of particular interest in studying the turbulent cascade is the relative magnitude of the thermal damping, as compared to the damping due to ion-neutral friction. The thermal damping rate in the modified loaded regime is given by Equation \eqref{eq:dampAppend}. The damping rate due to ion-neutral friction follows from the expression derived for the loaded regime, Equation \eqref{eq:ionNeutralDampingRate}: in the modified regime, we have to replace $c_n$ with $c_{{\rm M}n}$, which is equivalent to setting $\chi=1$ in the expression for $\mathscr{G}_{\rm f,s}(\beta_{\rm M}/\chi, \theta)$, Equation~\eqref{eq:fancyG}.

Thus, the ratio of thermal damping to damping from ion-neutral friction in the modified regime is 
\begin{equation}
    \mathscr{R}_{\rm f,s} = (\chi^{-1}-1) \tau_{\rm c} \nu_{ni} \beta_{\rm M} \frac{\mathscr{F}_{\rm f,s}(\beta_{\rm M}, \theta)}{\mathscr{G}_{\rm f,s}(\beta_{\rm M}, \theta)}\,, 
    \label{eq:scriptR}
    \end{equation}
where $\mathscr{F}_{\rm f,s}$ and $\mathscr{G}_{\rm f,s}$ are given by Equations \eqref{eq:fancyF} and \eqref{eq:fancyG} respectively. For $\beta_{\rm M} \gg 1$, where the fast mode is continuous, we have $\beta_{\rm M} \mathscr{F}_{\rm f}/\mathscr{G}_{\rm f} \rightarrow  \beta_{\rm M}^2/\sin^2{\theta}$; for $\beta_{\rm M} \ll1$, the slow mode is continuous, and $\beta_{\rm M} \mathscr{F}_{\rm s}/\mathscr{G}_{\rm s} \rightarrow  \cot^2{\theta}$. We conclude that the magnitude of $\mathscr{R}_{\rm f,s}$ for the continuous mode (where it exists, see Figure~\ref{continuityImage}) is, generally, well described by the product $\tau_{\rm c}\nu_{ni}$; for large $\beta_{\rm M}$ and/or small $\theta$, it is further enhanced.

The ratio $\mathscr{R}$, evaluated at $\theta = \pi/4$ for the continuous mode, as well as the wavelengths of the decoupling and damping peak, $\lambda_{\rm dec}$ and $\lambda_{\rm peak}$, are given in Table 2 for different environments considered above. The left group of columns corresponds to our standard value of $c_{\rm LA}/c_n = 1.4$.  The right group shows that the thermal damping becomes much more important in a higher-$\beta$ regime. We conclude that the effect of thermal damping on the cascade at large scales is very important for the cloud envelope, somewhat relevant for the CNM and outer core environments, and probably unimportant for the inner core. In cases where the thermal damping is not important in the loaded regimes, it may still play a significant role in the independent cascade of neutrals, which occurs at smaller scales. The large difference in $\mathscr{R}$ between the cloud envelope and the outer core phases is largely attributable to the sharp drop-off in ionization fraction, assumed to occur between those densities.  $\mathscr{R}$ is large in the WNM, although we note that collisionless damping may play a significant role in the WNM as well, depending on the ionization fraction.

\vspace{.7cm}

\section{Turbulent cascade}
\label{cascade}

In order to assess the importance of the thermal damping mechanism, we include it in a simple model of the turbulent
cascade. We consider both a Kolmogorov and a Kraichnan cascade. A model equation for the steady-state spectral energy
density of waves with the amplitude damping rate $\Gamma$ is given in \citet{Ptuskin06}:
\begin{equation}
\frac{d}{dk} \left(\frac{k W}{\tau_{\rm nl}}\right) = -2\Gamma W.
\label{Ptuskineq}
\end{equation}
Here, $\Gamma(k) = -{\rm Im}~\omega(k)$ and $W(k)$ is the turbulent energy spectrum of magnetic field fluctuations. We
assume equipartition between the energy spectrum of magnetic fluctuations and the kinetic energy density of the turbulence,
such that
\begin{equation}
kW(k) = \frac12\rho v^2(k).
\label{eq:Weq}
\end{equation}
$\tau_{\rm nl}(W, k)$ is the characteristic timescale of the nonlinear cascade.  This is given in \citet{Ptuskin06} as
\begin{equation}
\tau_{\rm nl}^{\rm Kol} = C_{\rm Kol} \sqrt{\frac{\rho}{k^3W}}\,,
\label{eq:tauKolmogorov}
\end{equation}
and
\begin{equation}
\tau_{\rm nl}^{\rm Kr} = C_{\rm Kr} \frac{\rho c_{\rm LA}}{k^2 W}\,,
\label{eq:tauKraichnan}
\end{equation}
where $C_{\rm Kol} \approx 12$ and $C_{\rm Kr} \approx 1$.

\begin{figure}
\centering
\includegraphics[width=\columnwidth]{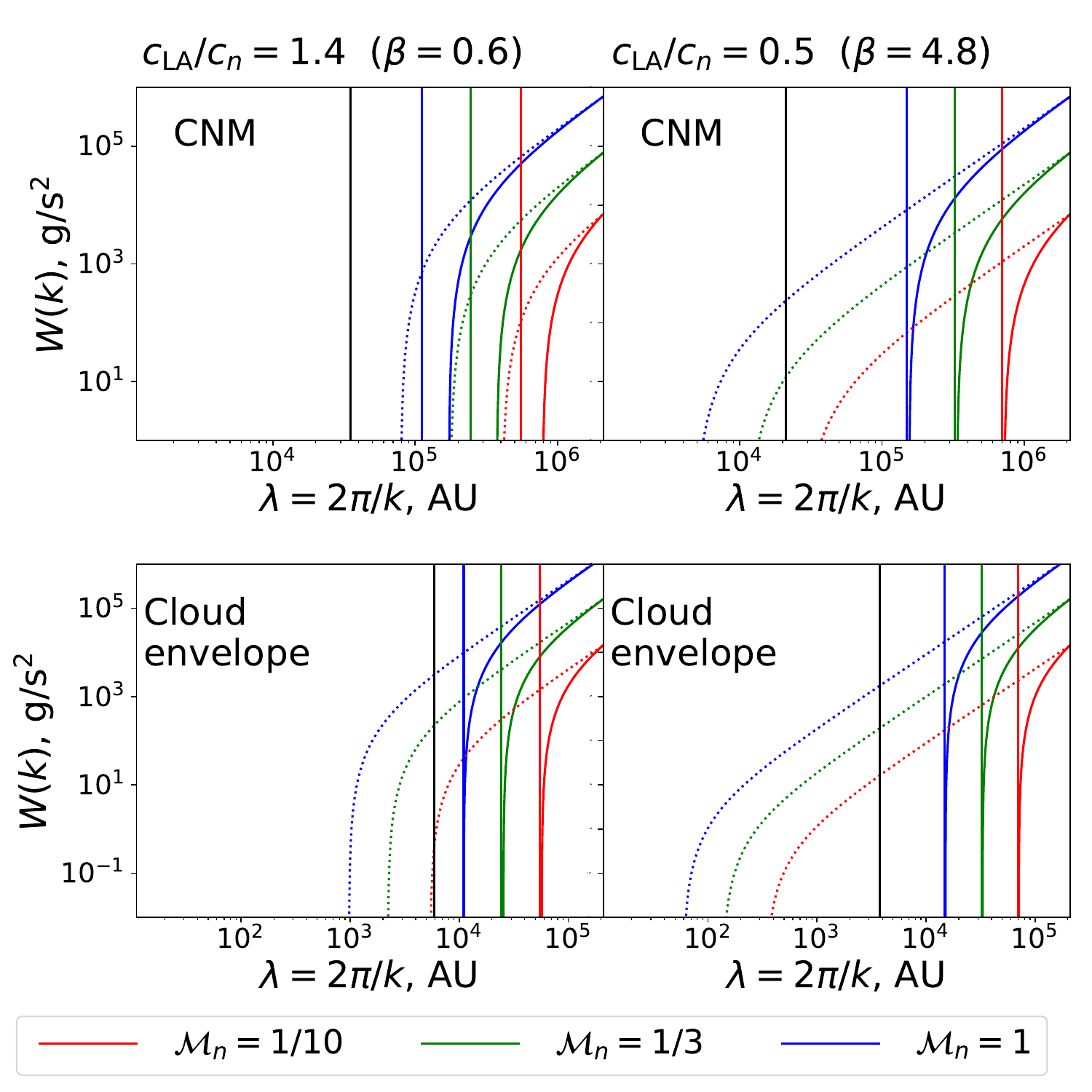}
\caption{The turbulent energy spectrum described by Equation \eqref{Ptuskineq}, with $\tau_{\rm nl}$ given by
Equation~\eqref{eq:tauKolmogorov} appropriate for a Kolmogorov cascade. Different colors correspond to different injection
velocities $v_0$, as shown in the legend in terms of the sonic Mach number $\mathcal{M}_n = v_0/c_n$.  Thermal damping is
included for the solid lines, but omitted for the dotted lines.  Turbulence is injected at a scale $\lambda_0$ of 1 pc for
the cloud envelope, and 10 pc for the CNM.  In the left panels, the magnetic field is chosen such that $c_{\rm LA} = 1.4 c_n$,
as in Figure~\ref{fourPanel}; in the right panels, $c_{\rm LA} = 0.5 c_n$. The colored vertical lines correspond to our
analytic estimate for the damping scale $\lambda_{\rm damp}$ provided in Equation \eqref{eq:KolmogorovDampingScale}. The
black vertical lines indicate the condition Re~$\omega = \tau_{\rm c}^{-1}$.} \label{kolmogorovPlot}
\end{figure}

\begin{figure}[htbp]
\centering
\includegraphics[width=\columnwidth]{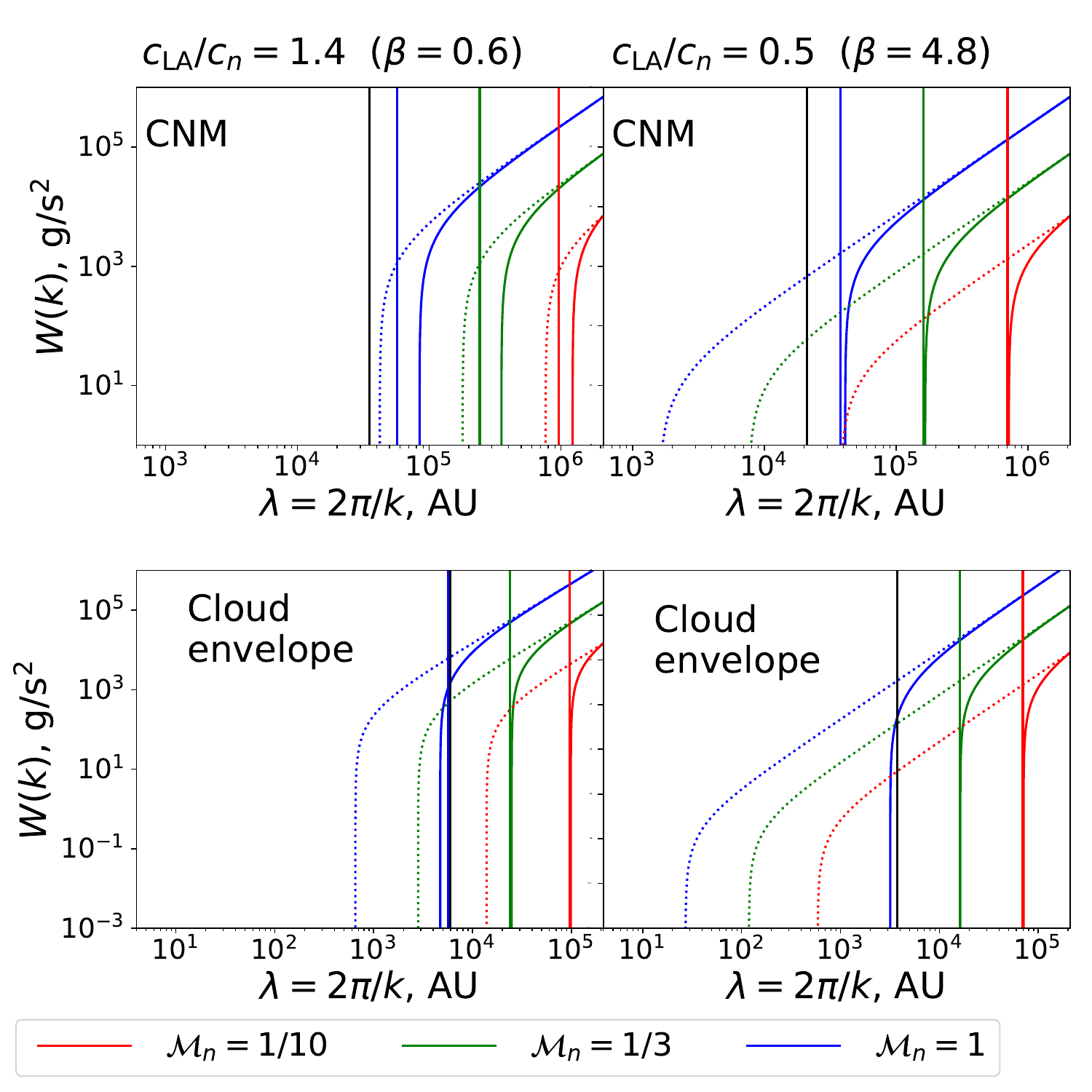}
\caption{Analogous to Figure \ref{kolmogorovPlot}, but $\tau_{\rm nl}$ was taken from Equation \eqref{eq:tauKraichnan}
appropriate for a Kraichnan cascade.}
\label{kraichnanPlot}
\end{figure}

Figure~\ref{kolmogorovPlot} shows the steady-state turbulent spectrum provided by Equation \eqref{Ptuskineq} with the
Kolmogorov cascade time given by Equation \eqref{eq:tauKolmogorov}.  We took $\Gamma(k)$ appropriate for the continuous mode
(depicted by the green line in Figure \ref{fourPanel}).  The left panels are for the same parameters as considered in Figure
\ref{fourPanel}. In the right panels, the magnetic field strength has been reduced by a factor of $\approx 2.8$, so that
$c_{\rm LA}/c_n=0.5$ (corresponding to $\beta = 4.8$). Different color curves correspond to different turbulent velocities
at the injection scale (10 pc for the CNM, and 1 pc for the cloud envelope).  In all cases, we have used the wave modes for
a propagation angle of $45^\circ$ as representative.  As shown in Section \ref{sec:cont}, it is only for a narrow range of
$\beta$ that there is a strong variation of the damping of the continuous mode with propagation direction.\footnote{Of the
four cases shown in Figures~\ref{kolmogorovPlot} and \ref{kraichnanPlot}, the damping strongly depends on $\theta$ only for
high $\beta$ and the CNM environment.}

It is evident that the thermal damping has a profound effect on the turbulent spectrum. Comparing the right and left panels
in Figure~\ref{kolmogorovPlot}, it is apparent that even a relatively modest decrease in the magnetic field (below the value
which we consider in our ``standard'' case depicted in Figure \ref{fourPanel}) allows the cascade to proceed to much smaller
$\lambda$ if the thermal damping is not considered.  On the contrary, when thermal damping is included, the cascade is cut
off at slightly {\it larger} $\lambda$ as the magnetic field is reduced.  This is because the restoring force at lower
magnetization is more due to gas pressure and is thus more heavily influenced by the cooling.

The black vertical lines in Figure~\ref{kolmogorovPlot} show the locations of the peak of the damping, $\lambda_{\rm peak}$,
approximated by the condition ${\rm Re}~\omega = \tau_{\rm c}^{-1}$ (and indicated in Figure~\ref{fourPanel} by the green vertical
line for the continuous mode). The turbulence is cut off near or slightly above these values of $\lambda$.  For stronger
turbulence at the injection scale, the cascade is damped at shorter $\lambda$, since $\tau_{\rm nl}$ is smaller for larger
$W$.

In the modified loaded regime $\omega \ll \tau_{\rm c}^{-1}$, the thermal damping is much stronger than the damping due to ion-neutral
friction. We substitute the damping rate $\Gamma(k)=Fk^2$, where $F$ is evaluated from Equation \eqref{eq:Req} at
$\theta=45^\circ$. Then, Equation \eqref{Ptuskineq} with $\tau_{\rm nl}$ from Equation \eqref{eq:tauKolmogorov} has the
solution
\begin{equation}
\frac{k^\frac{5}{3}W(k)}{k_0^\frac{5}{3}W_0} = 1 - \frac{\sqrt{2} \pi C_{\rm Kol} F}{\mathcal{M}_n c_n\lambda_0}
\left[\left(\frac{k}{k_0}\right)^\frac{4}{3} -1\right],
\label{eq:23}
\end{equation}
where $\mathcal{M}_n = v_0/c_{n}$ is the sonic Mach number, introduced via the relation $k_0W_0 =\frac12\rho v_0^2$, with
the subscript 0 referring to the value of the quantity at the injection scale. We can solve Equation \eqref{eq:23} for
$k_{\rm damp}$ such that $W(k_{\rm damp}) = 0$, which gives a damping wavelength $\lambda_{\rm damp} \equiv 2 \pi/k_{\rm
damp}$ of
\begin{equation}
\lambda_{\rm damp}^{\rm Kol} = \lambda_0 \left(1 + \frac{\mathcal{M}_n c_n\lambda_0}{\sqrt{2}\pi C_{\rm Kol}F}
\right)^{-\frac{3}{4}}.
\label{eq:KolmogorovDampingScale}
\end{equation}
These values for different $\mathcal{M}_n$ are shown by the vertical lines in Figure~\ref{kolmogorovPlot}.  

Figure \ref{kraichnanPlot} is the same as Figure \ref{kolmogorovPlot}, but $\tau_{\rm nl}$ is given by Equation
\eqref{eq:tauKraichnan}, appropriate for a Kraichnan cascade.  The differences are very minor.  Using Equation
\eqref{eq:Weq}, one can write the ratio of the cascade times given by Equations \eqref{eq:tauKolmogorov} and
\eqref{eq:tauKraichnan} as $\tau_{\rm nl}^{\rm Kol}/\tau_{\rm nl}^{\rm Kr} \approx 8.4 v(k)/c_{\rm LA}$.  This ratio is of
order unity for most of our examples, so the cutoff occurs at similar $k$.

Under the same assumptions as for the Kolmogorov case, by using $\tau_{\rm nl}^{\rm Kr}$ in Equation \eqref{Ptuskineq} we
find an analytic solution:
\begin{equation}
\frac{k^\frac{3}{2}W(k)}{k_0^\frac{3}{2}W_0} = 1- \frac{8\pi C_{\rm Kr} F }{3 \mathcal{M}_{\rm LA}^2c_{\rm LA}\lambda_0}
\left[\left(\frac{k}{k_0}\right)^\frac{3}{2}-1\right],
\end{equation}
where $\mathcal{M}_{\rm LA} = v_0/c_{\rm LA}$ is the Alfv{\'e}nic Mach number. Setting $W(k)$ to zero, we solve for the
damping wavelength
\begin{equation}
\lambda_{\rm damp}^{\rm Kr} = \lambda_0 \left(1 + \frac{3\mathcal{M}_{\rm LA}^2 c_{\rm LA}\lambda_0}{8\pi C_{\rm Kr}F}
\right)^{-\frac{2}{3}},
\label{eq:KraichnanDampingScale}
\end{equation}
shown by the vertical lines in Figure \ref{kraichnanPlot}. We point out that if $\lambda_{\rm damp}$ were significantly
smaller than  the wavelength  of peak damping $\lambda_{\rm peak}$, then the above analysis (assuming $\Gamma \propto k^2$)
would break down.  In this case the turbulence would continue to cascade to smaller wavelength (where $\Gamma$ is constant)
until it is eventually damped by viscosity. For the environments we consider, however, this would only happen if
$\mathcal{M}_n >1$, at which point our analysis is not valid anyway.

\section{MHD Simulations with cooling} \label{simulation}

\begin{figure}[htbp]
\centering
\includegraphics[width=0.98\columnwidth]{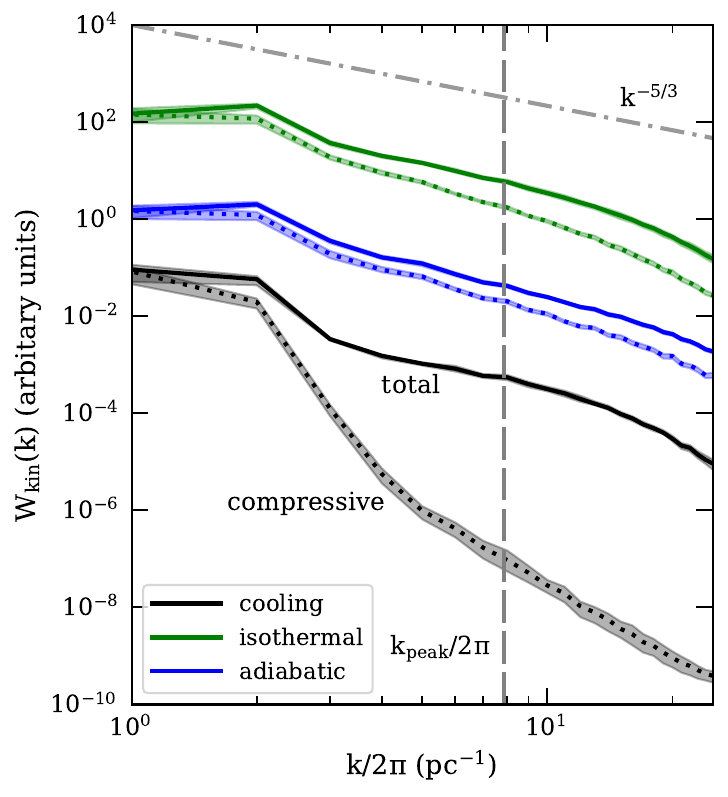}
\caption{Kinetic energy spectra $W_{\rm kin}(k)$ from MHD simulations with gas
heating and  cooling (black), as well as with isothermal (green) and adiabatic (blue) equations of state.
The vertical axis is in arbitrary units and the curves in each simulation
are shifted vertically for clarity.
In each case, the solid line shows the time-averaged spectrum at steady state
for the total kinetic energy and the dotted line represents the compressive modes.  The shaded area indicates the $1\sigma$ dispersion among the 
different time snapshots. The gray dash-dotted line indicates the slope
for the Kolmogorov spectrum of $k^{-5/3}$. The vertical gray dashed line
shows the location for the fastest damped mode at $k_{\rm peak}/2\pi=1/\lambda_{\rm peak}$ from the linear
theory. In the isothermal or adiabatic cases, the energy
spectrum of the compressive modes has a similar shape to the total
energy spectrum and only slightly less power. In the case with gas heating and cooling, the compressive mode is damped strongly, with a much steeper spectral slope and much less power.}
\label{munanPlot}
\end{figure}

In addition to the analytic approach followed in the previous sections, we also ran MHD
simulations with continuously driven turbulence. We solve the ideal MHD equations \citep[see, e.g.,][]{Stone2008},\footnote{In Equations (\ref{eq:momentum})--(\ref{eq:Bcurl}),
the $1/\sqrt{4\pi}$ pre-factor is absorbed in the unit of $\mathbf{B}$.}
\begin{equation}\label{eq:mass}
    \frac{\partial \rho}{\partial t} + \nabla\cdot(\rho \mathbf{v}) = 0,
\end{equation}
\begin{equation}\label{eq:momentum}
\frac{\partial \rho\mathbf{v}}{\partial t} 
    + \nabla\cdot(\rho \mathbf{vv} - \mathbf{BB})
    + \nabla\left(p + \frac{1}{2}B^2\right)
    = 0,
\end{equation}
\begin{equation}\label{eq:energy}
  \frac{\partial E}{\partial t} + \nabla \cdot \left[(p+E)\mathbf{v} - (\mathbf{B\cdot v})\mathbf{B}\right] =
  \Lambda_{\rm h} - \Lambda_{\rm c}\,,
\end{equation}
\begin{equation}\label{eq:eos}
  E =\frac{p}{\gamma - 1} + \frac{1}{2}\rho v^2 + \frac12B^2\,, 
\end{equation}
\begin{equation}\label{eq:Bcurl}
    \frac{\partial \mathbf{B}}{\partial t} 
    - \nabla \times (\mathbf{v} \times \mathbf{B}) = 0,
\end{equation}
where $\rho$ is the mass density of gas, $\mathbf{v}$ is the gas velocity, $p$ is the gas pressure, $\mathbf{B}$ is the magnetic field, $E$ is the total energy density, and $\Lambda_{\rm h}$ and $\Lambda_{\rm c}$ are the gas heating and cooling rates. The adiabatic index $\gamma$ is set to $5/3$.

The values of $\Lambda_{\rm h}$ and $\Lambda_{\rm c}$ in our simulations correspond to the CNM conditions, as described in Section \ref{sect:CNM}. Here we also include helium with abundance $x_{\rm He} = 0.1$. The mean molecular weight of the gas is then 1.4 accounting for helium.  The specific heat capacity per H atom at constant volume is given by $C_v^{\rm H} = 1.5 k_{\rm B} (1 + x_\mathrm{He})$,
ignoring the minor contribution from metals and electrons. 
In addition, we run two simulations with an isothermal and an adiabatic equation of state. For the adiabatic simulation, $\Lambda_{\rm h}=\Lambda_{\rm c}=0$; for the isothermal simulation, Equations~(\ref{eq:energy}) and (\ref{eq:eos}) are replaced by an isothermal equation of state, $p/\rho=c_n^2/\gamma$, and the sound speed $c_n$ is calculated at the equilibrium CNM temperature of 60~K (see Section \ref{sect:CNM}).     

We perform the simulations using the MHD code Athena++ \citep{Stone20}. The simulation set-up is very similar to the driving turbulence simulations in \citet{Gong20}, and we refer the readers to that paper for more details.
We adopt a Cartesian coordinate system with periodic boundary conditions on all sides. We use the Harten-Lax-van Leer-Discontinuities (HLLD) Riemann solver \citep{MK05}, with the third-order Runge–Kutta integrator and third-order spatial reconstruction. For each simulation,
turbulence was driven with an energy spectrum of $E(k)\propto k^{-3}$ at integer values of $[\tilde k_x, \tilde k_y, \tilde k_z]$ such that $ 0 < \tilde k = \sqrt{\tilde k_x^2 + \tilde k_y^2 + \tilde k_z^2} < 2$.  $\tilde k \equiv k/k_L$, where $k_L$ is the value corresponding to the box size.  The turbulence is driven with a fully compressive velocity field, but there is a small solenoidal component in the driving due to the finite forcing correlation time \citep{Grete18}. The simulations are run with a box-size of $L_x=L_y=L_z=1~\mathrm{pc}$, and a resolution of $N_x=N_y=N_z=180$. For the fast mode, with $\theta = 45^\circ$, thermal damping is peaked at $\lambda_{\rm peak} = 0.13$ pc, and we assume this to be the scale at which the damping effect is strongest.  With this resolution, the scale $\lambda_{\rm peak}$ is well resolved by 
23 cells.
In the isothermal simulation, we adopt the sound speed at the equilibrium temperature, 0.62 km/s. In the adiabatic simulation, we use the adiabatic index of 5/3 and start with the gas at the equilibrium temperature. The gas heats up gradually (the temperature increases by about 10\% at the end of the simulation) as kinetic energy is 
injected by turbulence driving. 
In the simulation, we include the heating and cooling as explicit source terms in the energy equation.
We limit the simulation time step so that the gas temperature in each cell changes less than 10\% during each time step. 

 We run the simulations for about 1000 Myr, more than 10 turbulence crossing times, and analyse the results at steady state between 600 -- 1000 Myr. The steady state sonic turbulent Mach number $\mathcal{M}_n$ at a scale of 1 pc is 0.02--0.08.  We pick this low value so as to avoid the introduction of non-linear effects which might complicate the analysis.  We start with a plasma $\beta=2$ in all simulations, and its value does not vary significantly in these cases due to the weak turbulence. 
 We experimented with a lower resolution of $N_x=N_y=N_z=100$, a higher steady-state Mach number around 0.2, and a higher $\beta$ around 10, and found very similar results to the simulations shown here.

The energy spectra of the turbulence from the simulations are shown in Figure \ref{munanPlot}. The lines are vertically offset for clarity.  In the isothermal and adiabatic simulations, the energy spectrum in the compressive modes are similar to the total energy spectrum, with only slightly less power and slightly steeper slopes. 
With realistic cooling, the
compressive modes decay much more rapidly with $k$ than the solenoidal ones, resulting in a much steeper spectrum and much less power in the compressive modes. 

The agreement with the analytic model is at this stage only qualitative.  Due to the limited numerical resolution, we cannot precisely describe the turbulent spectrum.  At
high $k$, the spectrum is affected by numerical damping, and at low $k$ the cascade is not fully developed. In the analytic
case, the damping cuts off the spectrum sharply at a particular scale.  In the simulation, there is not a sharp cutoff, but
rather a substantially steeper slope over a wide range of $k$. Despite of these limitations, however, the simulations clearly show that the compressive modes of the turbulence are damped efficiently by gas cooling.

\section{Implications}
\label{sect:impl}
The detailed implications of thermal damping of compressive modes require more realistic simulations which will better determine the magnitude of the effect on the turbulent cascade in realistic conditions.  However, in this section we outline a few different possible area in which thermal damping may be important.
\subsection{CR propagation}
The CR propagation in the ISM is often modelled as a diffusion process \citep{Strong98, Strong07}.  The scattering which determines the diffusion coefficient primarily arises from perturbations of the magnetic field which are resonant with the gyroradius of CRs \citep{Berezinskii1984,Strong07}. 
It was shown in \citet{Yan04} that in realistic ISM conditions, fast modes dominate the scattering of CRs. Here we demonstrate that, depending on the plasma $\beta$, such modes are strongly damped in the WNM and molecular cloud environments, which thus can significantly affect CR propagation.
\subsection{Dust growth}
Dust coagulation in molecular clouds is sensitive to the degree of turbulence on very small scales \citep{Voelk80, YLD04, Ormel07}, and can begin at densities as low as $10^4$ cm$^{-3}$ \citep{Hirashita09, Silsbee20}.  If the turbulence is damped at larger
scales, the evolution of the dust population may proceed more slowly.  This can have implications for the gas freeze-out on the surface of dust grains in dark clouds (which also occurs around densities of $10^4$ cm$^{-3}$).  Additionally, although the volume filling factor of such gas is small, coagulation at somewhat lower densities ($10^3$ cm$^{-3}$) can affect the grain size distribution in the ISM as a whole, if such material is mixed back into the more diffuse phases \citep[see, e.g.,][]{Hirashita19}.
\subsection{Observations of small-scale motions}
Although small-scale motions cannot typically be directly resolved in observations, there are strong indications that turbulence on small scales does exist.  This follows both from scintillation observations of electron density fluctuations in the diffuse ISM \citep{Armstrong95} and observations showing different velocity dispersions of ionized and neutral tracers in molecular clouds \citep{Li08, Pineda21}.  A quantitative understanding of the relevant damping processes is therefore essential for the development of a theoretical framework which could explain these observations.

\section{Conclusion}
\label{sect:conclusion}

In this work we introduced and analyzed the effect of thermal damping, associated with radiative cooling of gas, on compressive MHD turbulence in the ISM.  For a particular wave mode, the scale at which the damping is maximized depends on the cooling timescale and the plasma $\beta$.  In some environments the damping peaks at scales hundreds of times the ambipolar damping scale --
the characteristic wavelength at which the ions and neutrals decouple.  We confirm, using a simple analytic model of the
turbulent cascade as well as MHD simulations including cooling, that thermal damping has a substantial effect on the
turbulent spectrum of compressive waves for parameters appropriate for several characteristic phases of the ISM.  We find, for typical turbulent
velocities, that the turbulent cascade is cut off at a scale comparable to the scale at which the thermal damping peaks.  The spatial scale of this peak varies from environment to environment.  In typical WNM
conditions, the peak occurs at scales of around 100 pc, whereas in the CNM or in low-density molecular gas, the effect peaks
around 0.1 pc.  At densities greater than $10^5$ cm$^{-3}$, typical for the inner parts of molecular cloud cores, the peak becomes on the order of a milliparsec.  

We derived Equation~\eqref{eq:scriptR} for the strength $\mathscr{R}$ of the thermal damping relative to damping from ion-neutral friction. Its magnitude is determined by the product $\tau_{\rm c}\nu_{ni}$ of the cooling timescale and the neutral-ion collision frequency. For the conditions we considered, thermal damping is most efficient in low-density molecular gas, but can also play a significant role in the CNM and outer parts of a prestellar core, (as well as in the WNM, although collisionless damping may be important there as well).

This effect is important for a number of reasons. An open question in the study of ISM turbulence is how the turbulent energy is transferred from large scales, where ions and neutrals move together, to small scales, where they move separately \citep[see][]{Tilley11, Burkhart15}. This process may involve compressive modes, since Alfv{\'e}n waves in low-ionization media have a relatively wide range of $k$ in which they cannot propagate.  For this reason, it is important to have a complete understanding of the mechanisms which impact the survival of compressive waves.  Additionally, the damping described in this paper has potential implications for dust evolution in the ISM, propagation of CRs, and the interpretation of observed line widths of ionized versus neutral gas tracers in molecular clouds.
\par
A.V.I. acknowledges support by the Russian Science Foundation (project 18-12- 00351).

\appendix
\section{Appendix A: Rate of Thermal Damping}
\label{app:modifiedDamping}

Thermal damping generally dominates in the modified loaded regime $\omega \tau_{\rm c} \ll 1$. This allows us to ignore ion-neutral
friction and derive an analytic formula for the damping rate, considering the limit of small ionization fraction for
simplicity.  The dispersion relation of fast and slow modes in the loaded regimes (on the right from the decoupling gap in Figure \ref{fourPanel}) is generally given by the standard expression \citep{Landau60} where $c_{\rm A}$ is replaced, according to Equation \eqref{eq:EOS}, with $c_{\rm LA}$, while $c_i$ is replaced with $c_n f(\omega)$.
\begin{equation}
    \omega^4 - \omega^2 k^2 c_n^2 f(\omega) - \omega^2 k^2 c_{\rm LA}^2 + k^4 \cos^2{\theta} c_n^2 c_{\rm LA}^2 f(\omega) = 0.
    \label{eq:disp5}
\end{equation}
In the modified loaded regime, where $\omega \tau_{\rm c} \ll 1$, we can approximate Equation \eqref{eq:fdef} as $f(\omega) = \chi(1-\eta \omega \tau_{\rm c})$, where 
 \begin{equation}
 \eta = \chi^{-1} - 1
 \end{equation}
 is a positive number.  In this case, Equation \eqref{eq:disp5} becomes
\begin{equation}
\omega^4 - \omega^2k^2 \left[c_{\rm LA}^2 + c_{{\rm M}n}^2(1-i\eta \omega \tau_{\rm c})\right] + k^4 c_{\rm LA}^2 c_{{\rm M}n}^2(1-
i\eta \omega \tau_{\rm c}) \cos^2{\theta} = 0,
\label{eq:disp5modified}
\end{equation}
where $c_{{\rm M}n}$ is given by Equation \eqref{eq:cmn}.

We can present the solution for the modified loaded (fast or slow) mode as
 \begin{equation}
 \omega = kc_{\rm Mf,s} +i\:{\rm Im}~\omega,
 \label{eq:omegaPert}
 \end{equation}
where $c_{\rm Mf,s}$ is the fast or slow mode speed, given by Equation \eqref{eq:fastMode} with $c_{\rm A}$ replaced with $c_{\rm
LA}$, and $c_i$ replaced with $c_{{\rm M}n}$. Substituting Equation \eqref{eq:omegaPert} into Equation \eqref{eq:disp5modified}, and
keeping only first-order terms in ${\rm Im}~\omega$ gives
\begin{equation}
 -{\rm Im}~\omega = \frac{\eta \tau_{\rm c} c_{{\rm M}n}^2}{2} \left(\frac{c_{\rm Mf,s}^2  - c_{\rm LA}^2
 \cos^2{\theta}}{2 c_{\rm Mf,s}^2 - c_{\rm LA}^2 - c_{{\rm M}n}^2}\right)k^2 \equiv F_{\rm f,s}k^2.
\end{equation}
After some manipulation, we obtain
\begin{equation}
F_{\rm f,s} = \frac{\eta \tau_{\rm c} c_{{\rm M}n}^2}{4} \,\mathscr{F_{\rm f, s}}(\beta_{\rm M}, \theta),
\label{eq:Req}
\end{equation}
where $\beta_{\rm M} = c_{{\rm M}n}^2/c_{\rm LA}^2 \equiv \gamma \chi \beta/2$, and
\begin{equation}
    \mathscr{F}_{\rm f,s}(\beta_{\rm M}, \theta) = 1\mp \frac{\cos{2 \theta} - \beta_{\rm M}}{\sqrt{(\cos{2\theta} - \beta_{\rm M})^2 +
    \sin^2{2\theta}}}\,.
    \label{eq:fancyF}
\end{equation}
We note that $\mathscr{F}_{\rm s} +\mathscr{F}_{\rm f}=2$.

\section{Appendix B: Loaded regime and decoupling gap}
\label{App_gap}

A general dispersion relation given by Equations~\eqref{eq20}--\eqref{eq21} can be substantially simplified for wave modes
sustained in the loaded regime and in the decoupling gap (see Section~\ref{sect:effect}). For this, we require the ratio of
the ion-to-neutral densities, $\rho_i/\rho_n$, to be a sufficiently small number, so that the relative width of the
decoupling gap is large, $\sim \sqrt{\rho_n/\rho_i}$ \citep[e.g.,][]{Kulsrud69}. As damping near the decoupling gap is
dominated by ion-neutral friction, here we can neglect thermal damping and set $f(\omega)=1$.

Let us identically rewrite Equation~\eqref{eq20a}, by replacing in the first term $\Delta_n$ with $\Delta_i$ and subtracting
from the resulting equation the corresponding difference $\propto(\Delta_i-\Delta_n)$. Equation~\eqref{eq21} for neutrals is
rewritten in a similar way, by replacing $\Delta_n^\|$ with $\Delta_i^\|$ in the first term, and $\Delta_n$ with $\Delta_i$
in the second term. Then, excluding $\Delta_i-\Delta_n$ and $\Delta_i^\|-\Delta_n^\|$ from the resulting four equations, we
obtain:
\begin{equation}\label{disp1}
\left[\omega^2-k^2(c_{\rm A}^2+c_i^2)+\frac{i\omega\nu_{in}}{\omega^2+i\omega\nu_{ni}-k^2c_n^2}\,(\omega^2-k^2c_n^2)\right]
\Delta_i+k^2c_{\rm A}^2\,\Delta_i^\|=0,
\end{equation}
\begin{equation}\label{disp2}
\omega^2\,\Delta_i^\|-\frac{i\omega\nu_{ni}}{\omega^2+i\omega\nu_{ni}-k^2c_n^2}\,k_z^2c_n^2\,\Delta_i=0.
\end{equation}
We are interested in wave modes with $|\omega|\sim\nu_{ni}$, corresponding to the right edge of the decoupling gap (see
below), and therefore in deriving Equation~\eqref{disp2}, we neglected terms $\propto \nu_{ni}/\nu_{in} =
\rho_i/\rho_n$, and terms $\propto \omega/\nu_{in}\equiv (\rho_i/\rho_n)(\omega/\nu_{ni})$. For the same
reason, the third term in the brackets in Equation~\eqref{disp1} is estimated as $\sim(\nu_{in}/\nu_{ni})\omega^2$,
and therefore the first term $\omega^2$ can be omitted. We can also neglect $c_i^2$ next to $c_{\rm A}^2$, because $c_{\rm
LA}\equiv c_{\rm A}\sqrt{\rho_i/\rho_n}$ in our analysis is assumed to be comparable to $c_n\sim c_i$. Substituting $\nu_{in}=
(\rho_n/\rho_i)\nu_{ni}$, after some manipulation we obtain the following dispersion relation:
\begin{equation}\label{disp3}
k^2c_{\rm LA}^2\omega(\omega^2-k^2c_n^2)=i\nu_{ni}(\omega^2-k^2c_{\rm Lf}^2)(\omega^2-k^2c_{\rm Ls}^2),
\end{equation}
where $c_{\rm Lf,s}^2$ are given by Equation~\eqref{eq:fastMode} with $c_{\rm A}$ replaced by $c_{\rm LA}$, and $c_i$ replaced by
$c_n$.

Equation~\eqref{disp3} has an intuitive structure. For sufficiently small $k$, representing wavelengths $\lambda$ to the
right of the decoupling gap (see Section~\ref{sect:effect} and Figure~\ref{fourPanel}), the LHS is small (since $\omega
\ll \nu_{ni}$) and thus we recover the loaded fast and slow modes, Re~$\omega\approx kc_{\rm Lf,s}$ and 
\begin{equation}
-{\rm Im}~\omega = \frac{c_{\rm LA}^2}{2\nu_{ni}} \frac{|c_{\rm Lf, s}^2 - c_n^2|}{c_{\rm Lf}^2-c_{\rm Ls}^2}k^2 \equiv \frac{c_{\rm LA}^2}{4\nu_{\rm ni}} \mathscr{G}_{\rm f,s}(\beta_{\rm M}/\chi, \theta)k^2,
\label{eq:ionNeutralDampingRate}
\end{equation} 
where 
\begin{equation}
\mathscr{G}_{\rm f,s}(\beta_{\rm M}/\chi, \theta) =  1\pm \frac{1-\beta_{\rm M}/\chi}{\sqrt{\left(1-\beta_{\rm M}/\chi\right)^2 + 4(\beta_{\rm M}/\chi)\sin^2{\theta}}}\,.
\label{eq:fancyG}
\end{equation}
Note that the effect of ion-neutral friction in the modified loaded regime is obtained by following the same derivation for $f(\omega)=\chi$: this yields Re~$\omega\approx kc_{\rm Mf,s}$ and ${\rm Im}~\omega$ given by Equation~\eqref{eq:ionNeutralDampingRate} with $c_{\rm Lf,s}\to c_{\rm Mf,s}$ and $c_n\to c_{{\rm M}n}$, which is equivalent to $\chi\to1$ in Equation~\eqref{eq:fancyG}.
For large $k$, corresponding to the decoupling gap, we recover the neutral sound mode
with Re~$\omega\approx kc_n$ and $-{\rm Im}~\omega\propto \nu_{ni}$. Equation \eqref{disp3} is not valid on the left
side of the decoupling gap, because the assumption we made that $\omega \ll \nu_{in}$ breaks down in this regime.

In order to obtain Figure~\ref{continuityImage}, we introduce dimensionless wave speed $\tilde c$, fast and slow speeds
$\tilde c_{\rm Lf,s}$, and wavelength $\tilde \lambda$, defined as
\begin{equation}\label{normalized}
\tilde c=\frac{\omega}{kc_n}, \quad \tilde c_{\rm Lf,s} = \frac{c_{\rm Lf,s}}{c_n} \quad {\rm and}  \quad
\tilde \lambda= \frac{c_n}{c_{\rm LA}}\frac{\nu_{ni}}{kc_{\rm LA}}\,.
\end{equation}
Then Equation~\eqref{disp3} becomes
\begin{equation}\label{disp4}
\tilde c(\tilde c^2-1)=i\tilde\lambda\left(\tilde c^2-\tilde c_{\rm Lf}^2\right)
\left(\tilde c^2-\tilde c_{\rm Ls}^2 \right).
\end{equation}
A condition for the neutral sound mode to switch between the loaded fast and slow modes (depicted by the solid line in
Figure~\ref{continuityImage}) is easily derived by substituting $\tilde c=-1/\tilde c_*$ in Equation~\eqref{disp4}. This
transforms the fast mode into the conjugate slow mode (with the real part of the opposite sign) and vice versa. The
resulting equation for $\tilde c_*$ is then reduced to the form of Equation~\eqref{disp4} by setting $c_{\rm Lf}^2c_{\rm
Ls}^2=c_n^4$, which yields the sought condition. Using Equation~\eqref{eq:fastMode} for $c_{\rm Lf,s}^2$, we obtain that the
switch between the fast and slow modes occurs at
\begin{equation}\label{switch}
\frac{c_n}{c_{\rm LA}}=\cos\theta.
\end{equation}

We now can derive an approximate condition for the right edge of the decoupling gap. Substituting $x = i \tilde c$
transforms Equation \eqref{disp4} into an equation with all real coefficients. In the case that $c_{\rm Lf} \gg c_{\rm Ls}$,
we can find approximate solutions for the loaded fast mode by discarding the linear and constant terms in the transformed
equation, and for the loaded slow mode -- by discarding the 3rd- and 4th-order terms. This results in the following
equations:
\begin{eqnarray}
{\rm Loaded~fast:}&&\quad x^2 - \frac{x}{\tilde \lambda} + \tilde c_{\rm Lf}^2 + \tilde c_{\rm Ls}^2 = 0, \label{eq:Quadratic1}\\
{\rm Loaded~slow:}&&\quad \left(\tilde c_{\rm Lf}^2 + \tilde c_{\rm Ls}^2\right)x^2 - \frac{x}{\tilde \lambda} + \tilde
c_{\rm Lf}^2 \tilde c_{\rm Ls}^2 = 0.\label{eq:Quadratic2}
\end{eqnarray}
We keep in mind that if Equation~\eqref{eq:Quadratic1} or \eqref{eq:Quadratic2} has a pair of real solutions, then the
corresponding mode has no real solutions (i.e. the mode is discontinuous). For $c_n \gg c_{\rm LA}$, we expect the loaded slow mode to be cut off at the
right edge of the decoupling gap and the loaded fast mode to be continuous with the neutral sound mode (see Section
\ref{sec:cont}). Hence, by setting the discriminant of Equation~\eqref{eq:Quadratic2} to zero, we can solve for the right
edge in this limit: $k = 2 (\nu_{ni}/c_{\rm LA})\cos{\theta}$. Conversely, for $c_n \ll c_{\rm LA}$ we expect the loaded
fast mode to be cut off; setting the discriminant of Equation~\eqref{eq:Quadratic1} to zero yields $k = 2 \nu_{ni}/c_{\rm LA}$.

A condition that the continuous mode ceases to exist (depicted by the dashed lines in Figure~\ref{continuityImage}) is
equivalent to that Equation~\eqref{disp4} written for $x=i\tilde c$ has four real roots at $0<x<\infty$. In principle, this
condition could be derived analytically, using Sturm's theorem for the number of real roots of a polynomial, but we found it
more convenient to compute these boundaries numerically.

\bibliographystyle{apj}
\bibliography{Paper}

\end{document}